# Antigravitation, Dark Energy, Dark Matter – Alternative Solution

Boris V. Alexeev
Moscow Academy of Fine Chemical Technology (MITHT)
Prospekt Vernadskogo, 86, Moscow 119570, Russia
B.Alexeev@ru.net

**Abstract**

Collisional damping of gravitational waves in the Newtonian matter is investigated. The generalized theory of Landau damping is applied to the gravitational physical systems in the context of the plasma gravitational analogy. The existence of gravitational window is discovered where the waves of gravitating matter are expanding without damping.

## 1. Elementary introduction in the basic principles of the Generalized Boltzmann Physical Kinetics.

My aim consists in the construction of plasma – gravitational analogy using the generalized theory of Landau damping and the generalized Boltzmann physical kinetics developed by me.

The collisionless damping of electron plasma waves was predicted by Landau in 1946 [1, 2] and later was confirmed experimentally. Landau damping plays a significant role in many electronics experiments and belongs to the most well known phenomenon in statistical physics of ionized gases.

The physical origin of the collisionless Landau wave damping is simple. Really, if individual electron of mass $m_e$ moves in the periodic electric field, this electron can diminish its energy (electron velocity larger than phase velocity of wave) or receive additional energy from the wave (electron velocity less than phase velocity of wave). Then the total energy balance for a swarm of electrons depends on quantity of "cold" and "hot" electrons. For the Maxwellian distribution function, the quantity of "cold" electrons is more than quantity of "hot" electrons. This fact leads to, so-called, the collisionless Landau damping of the electric field perturbation.

In spite of transparent physical sense, the effect of Landau damping has continued to be of great interest to theorist as well. Much of this interest is connected with counterintuitive nature of result itself coupled with the rather abstruse mathematical nature of Landau's original derivation (including so-called Landau's rule of complex integral calculation). Moreover for these reason there were publications containing some controversy over the reality of the phenomenon (see for example, [3, 4]).

In paper [5] the difficulties originated by Landau's derivation were clarified. The mentioned consideration leads to another solution of Vlasov - Landau equation, these ones are in agreement with data of experiments. The problem solved in this article consists in investigation of Landau damping from the positions of Generalized Boltzmann Physical Kinetics [6 - 8]. The influence of the particle collisions is taken into account.

The results obtained in [5] are based on the classical Boltzmann equation (BE) written for collisionless case and can be derived as asymptotic solution of the generalized Boltzmann equation. Basic principles of non-local physics on the whole and the generalized Boltzmann physical kinetics in particular are formulated in [5 – 8]. Nevertheless I deliver here some main ideas and deductions of the generalized Boltzmann physical kinetics. For simplicity in introduction the fundamental methodic aspects are considered from the qualitative standpoint of view avoiding excessively cumbersome formulas. A rigorous description is found, for example, in the monograph [8].

In 1872 L Boltzmann published his famous kinetic equation for the one-particle distribution function $f(\mathbf{r},\mathbf{v},t)$. He expressed the equation in the form

$$Df/Dt = J^{st}(f), \qquad (1.1)$$

where $J^{st}$ is the collision integral, and $\dfrac{D}{Dt} = \dfrac{\partial}{\partial t} + \mathbf{v}\cdot\dfrac{\partial}{\partial \mathbf{r}} + \mathbf{F}\cdot\dfrac{\partial}{\partial \mathbf{v}}$ is the substantial (particle) derivative, $\mathbf{v}$ and $\mathbf{r}$ being the velocity and radius vector of the particle, respectively. Equation (1.1) governs the transport processes in a one-component gas, which is sufficiently rarefied that only binary collisions between particles are of importance and valid only for two character scales, connected with the hydrodynamic time-scale and the time-scale between particle collisions. While we are not concerned here with the explicit form of the collision integral, note that it should satisfy conservation laws of point-like particles in binary collisions. Integrals of the distribution function (i.e. its moments) determine the macroscopic hydrodynamic characteristics of the system, in particular the number density of particles $n$ and the temperature $T$. The Boltzmann equation is not of course as simple as its symbolic form above might suggest, and it is in only a few special cases that it is amenable to a solution. One example is that of a Maxwellian distribution in a locally, thermodynamically equilibrium gas in the event when no external forces are present. In this case the equality $J^{st} = 0$ and $f = f_0$ is met, giving the Maxwellian distribution function $f_0$.

A weak point of the classical Boltzmann kinetic theory is the way it treats the dynamic properties of interacting particles. On the one hand, as the so-called "physical" derivation of the BE suggests, Boltzmann particles are treated as material points; on the other hand, the collision integral in the BE brings into existence the cross sections for collisions between particles. A rigorous approach to the derivation of the kinetic equation for $f$ (noted in following as $KE_f$) is based on the hierarchy of the Bogolyubov-Born-Green-Kirkwood-Yvon (BBGKY) [8] equations.

A $KE_f$ obtained by the multi-scale method turns into the BE if one ignores the change of the distribution function (DF) over a time of the order of the collision time (or, equivalently, over a length of the order of the particle interaction radius). It is important to note [8] that accounting for the third of the scales mentioned above leads (*prior* to introducing any approximation destined to break the Bogolyubov chain) to additional terms, generally of the same order of magnitude, appear in the BE.

If the correlation functions is used to derive $KE_f$ from the BBGKY equations, then the passage to the BE means the neglect of non-local and time delay effects.

Given the above difficulties of the Boltzmann kinetic theory, the following clearly inter related questions arise:

First, what is a physically infinitesimal volume and how does its introduction (and, as the consequence, the unavoidable smoothing out of the DF) affect the kinetic equation? This question can be formulated in (from the first glance) the paradox form – what is the size of the point in the physical system?

And second, how does a systematic account for the proper diameter of the particle in the derivation of the $KE_f$ affect the Boltzmann equation?

In the theory developed here I shall refer to the corresponding $KE_f$ as the GBE. The derivation of the GBE and the applications of GBE are presented, in particular, in [8]. Accordingly, our purpose is first to explain the essence of the physical generalization of the BE.

Let a particle of finite radius be characterized, as before, by the position vector **r** and velocity **v** of its center of mass at some instant of time $t$. Let us introduce physically small volume (**PhSV**) as element of measurement of macroscopic characteristics of physical system for a point containing in this **PhSV**. We should hope that **PhSV** contains sufficient particles $N_{ph}$ for statistical description of the system. In other words all investigating physical system is covering by a net of physically small volumes.

Every **PhSV** contains entire quantity of point-like Boltzmann particles, and *the same DF $f$ is prescribed for whole **PhSV** in Boltzmann physical kinetics*. Therefore Boltzmann particles are fully "packed" in the reference volume.

Let us consider two adjoining physically small volumes **PhSV$_1$** and **PhSV$_2$**. We have on principle another situation for the particles of finite size moving in physical small volumes which are open thermodynamic systems.

Non- local effects take places. Namely:

The fact that center of mass of a particle is in **PhSV$_1$** does not mean that entire particle is there. In other words, at any given point in time there are always particles, which are partly inside and partly outside of the reference volume.

*Moreover the particles starting after the last collision near the boundary between two mentioned volumes can change the distribution functions in the neighboring volume. The adjusting of the particles dynamic characteristics for translational degrees of freedom takes several collisions. As result we have in the definite sense "the Knudsen layer" between these volumes. This fact unavoidably leads to fluctuations in mass and hence in other hydrodynamic quantities.*

*Existence of such "Knudsen layers" is not connected with the choice of space nets and fully defined by the reduced description for ensemble of particles of finite diameters in the frame of conception of physically small volumes, therefore – with the chosen method of measurement.*

The corresponding situation is typical for the theoretical physics – we could remind about the role of probe charge in electrostatics or probe circuit in the physics of magnetic effects.

Suppose that DF $f$ corresponds to **PhSV$_1$** and DF $f - \Delta f$ is connected with **PhSV$_2$** for Boltzmann particles. In the boundary area in the first approximation, fluctuations will be proportional to the mean free path (or, equivalently, to the mean time between the collisions). Then for **PhSV** the correction for DF should be introduced as

$$f^a = f - \tau\, Df/Dt \qquad (1.2)$$

in the left hand side of classical BE describing the translation of DF in phase space. As the result

$$Df^a/Dt = J^B, \qquad (1.3)$$

where $J^B$ is the Boltzmann local collision integral.

*Important to notice that it is only qualitative explanation of GBE derivation obtained earlier (see for example [8]) by different strict methods from the BBGKY – chain of kinetic equations.*

The structure of the $KE_f$ is generally as follows

$$\frac{Df}{Dt} = J^B + J^{nonlocal}, \qquad (1.4)$$

where $J^{nonlocal}$ is the non-local integral term incorporating the time delay effect. The generalized Boltzmann physical kinetics, in essence, involves a local approximation

$$J^{nonlocal} = \frac{D}{Dt}\left(\tau \frac{Df}{Dt}\right) \tag{1.5}$$

for the second collision integral, here $\tau$ being the mean time *between* the particle collisions. We can draw here an analogy with the Bhatnagar - Gross - Krook (BGK) approximation for $J^B$,

$$J^B = \frac{f_0 - f}{\tau}, \tag{1.6}$$

which popularity as a means to represent the Boltzmann collision integral is due to the huge simplifications it offers.

In other words – the local Boltzmann collision integral admits approximation via the BGK algebraic expression, but more complicated non-local integral can be expressed as differential form (1.5).

The ratio of the second to the first term on the right-hand side of Eq. (1.4) is given to an order of magnitude as $J^{nonlocal}/J^B \approx O(Kn^2)$ and at large Knudsen numbers (defining as ratio of mean free path of particles to the character hydrodynamic length) these terms become of the same order of magnitude. It would seem that at small Knudsen numbers answering to hydrodynamic description the contribution from the second term on the right-hand side of Eq. (1.4) is negligible.

*This is not the case, however.* When one goes over to the hydrodynamic approximation (by multiplying the kinetic equation by collision invariants and then integrating over velocities), the Boltzmann integral part vanishes, and the second term on the right-hand side of Eq. (1.4) gives a single-order contribution in the generalized Navier - Stokes description. Mathematically, we cannot neglect a term with a small parameter in front of the higher derivative. Physically, the appearing additional terms are due to viscosity and they correspond to the small-scale Kolmogorov turbulence [8]. The integral term $J^{nonlocal}$, thus, turns out to be important both at small and large Knudsen numbers in the theory of transport processes. Thus, $\tau Df/Dt$ is the distribution function fluctuation, and writing Eq. (1.3) without taking into account Eq. (1.2) makes the BE non-closed. From the viewpoint of the fluctuation theory, Boltzmann employed the simplest possible closure procedure $f^a = f$. For GBE the generalized H-theorem is proven [8].

Let us consider now some aspects of GBE application beginning with hydrodynamic aspects of the theory. There are two important points to be made here. First, the fluctuations will be proportional to the mean time *between* the collisions (rather than the collision time). This fact is rigorously established [8] and explained above. Therefore, in the first approximation, fluctuations will be proportional to the mean free path $\lambda$ (or, equivalently, to the mean time between the collisions). We can state that the number of particles in reference volume is proportional to cube of the character length $L$ of volume, the number of particles in the surface layer is proportional to $\lambda L^2$, and as result all effect of fluctuation can be estimated as ratio of two mentioned values or as $\lambda/L = Kn$.

Obviously the generalized hydrodynamic equations (GHE) will explicitly involve fluctuations proportional to $\tau$. For example, the continuity equation changes its form and will contain terms proportional to viscosity. On the other hand - and this is the second point to be made - if the reference volume extends over the whole cavity with the hard walls, then the classical conservation laws should be obeyed, and this is exactly what the monograph [8] proves. However, we will here attempt to "guess" the structure of the generalized continuity equation using the arguments outlined above. Neglecting fluctuations, the continuity equation should have the classical form with

$$\rho^a = \rho - \tau A, \ (\rho \mathbf{v}_0)^a = \rho \mathbf{v}_0 - \tau \mathbf{B},$$

where $\rho$ is density and $\mathbf{v}_0$ is hydrodynamic velocity. Strictly speaking, the factors $A$ and $\mathbf{B}$ can be obtained from the generalized kinetic equation, in our case, from the GBE. Still, we can guess their form without appeal to the $KE_f$.

Indeed, let us write the generalized continuity equation

$$\frac{\partial}{\partial t}(\rho - \tau A) + \frac{\partial}{\partial \mathbf{r}} \cdot (\rho \mathbf{v}_0 - \tau \mathbf{B}) = 0 \qquad (1.7)$$

in the dimensionless form, using $l$, the distance from the reference contour to the hard wall (see Fig. 1.1), as a length scale.

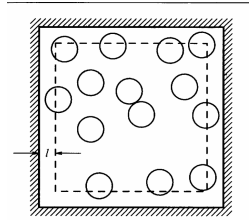

Fig. 1.1. Closed cavity and the reference contour containing particles of a finite diameter.

Then, instead of $\tau$, the (already dimensionless) quantities $A$ and $\mathbf{B}$ will have the Knudsen number $Kn_l = \lambda/l$ as a coefficient. In the limit $l \to 0, Kn_l \to \infty$ the contour embraces the entire cavity contained within hard walls, and there are no fluctuations on the walls. In other words, the classical equations of continuity and motion must be satisfied at the wall. Using hydrodynamic terminology, we note that the conditions $A = 0$, $\mathbf{B} = 0$ correspond to a laminar sub-layer in a turbulent flow. Now if a local Maxwellian distribution is assumed, then the generalized equation of continuity in the Euler approximation is written as

$$\frac{\partial}{\partial t}\left\{\rho - \tau\left[\frac{\partial \rho}{\partial t} + \frac{\partial}{\partial \mathbf{r}} \cdot (\rho \mathbf{v}_0)\right]\right\} + \frac{\partial}{\partial \mathbf{r}} \cdot \left\{\rho \mathbf{v}_0 - \tau\left[\frac{\partial}{\partial t}(\rho \mathbf{v}_0) + \frac{\partial}{\partial \mathbf{r}} \cdot \rho \mathbf{v}_0 \mathbf{v}_0 + \vec{I} \cdot \frac{\partial p}{\partial \mathbf{r}} - \rho \mathbf{a}\right]\right\} = 0, \quad (1.8)$$

where $\vec{I}$ is the unit tensor, $\mathbf{a}$ – the external force. In the hydrodynamic approximation, the mean time $\tau$ between the collisions is related to the dynamic viscosity $\mu$ by

$$\tau \, p = \Pi \mu, \qquad (1.9)$$

where the factor $\Pi$ depends on the choice of a collision model and is $\Pi = 0.8$ for the particular case of neutral gas comprising hard spheres [9]. The generalized hydrodynamic equations (GHE) of energy and motion are much more difficult to guess in this way, making the GBE indispensable.

Now several remarks of principal significance:
1. All fluctuations are found from the strict kinetic considerations and tabulated [8]. The appearing additional terms in GHE are due to viscosity and they correspond to the small-scale Kolmogorov turbulence. The neglect of formally small terms is equivalent, in particular, to dropping the (small-scale) Kolmogorov turbulence from consideration and is the origin of all principal difficulties in usual turbulent theory.

Fluctuations on the wall are equal to zero, from the physical point of view this fact corresponds to laminar sub-layer. Mathematically it leads to additional boundary conditions for GHE. Major difficulties arose when the question of existence and uniqueness of solutions of the Navier - Stokes equations was addressed. O A Ladyzhenskaya has shown for three-dimensional flows that under smooth initial conditions a unique solution is only possible over a finite time interval. Ladyzhenskaya even introduced a "correction" into the Navier - Stokes equations in order that its unique solvability could be proved (see discussion in [10]). GHE do not lead to these difficulties.

2. It would appear that in continuum mechanics the idea of discreteness can be abandoned altogether and the medium under study be considered as a continuum in the literal sense of the word. Such an approach is of course possible and indeed leads to Euler equations in hydrodynamics. But when the viscosity and thermal conductivity effects are to be included, a totally different situation arises. As is well known, the dynamical viscosity is proportional to the mean time $\tau$ between the particle collisions, and a continuum medium in the Euler model with $\tau = 0$ implies that neither viscosity nor thermal conductivity is possible.

3. The GBE for a plasma medium has the form [8]

$$\frac{Df_\alpha}{Dt} - \frac{D}{Dt}\left(\tau_\alpha \frac{Df_\alpha}{Dt}\right) = J_\alpha, \qquad (1.10)$$

where

$$\frac{D}{Dt} = \frac{\partial}{\partial t} + \mathbf{v}_\alpha \cdot \frac{\partial}{\partial \mathbf{r}} + \mathbf{F}_\alpha \cdot \frac{\partial}{\partial \mathbf{v}_\alpha} \qquad (1.11)$$

is the substantial (particle) derivative containing the self-consistent force $\mathbf{F}_\alpha$, $J_\alpha$ is the classical (Boltzmann) collision integral, and $\tau_\alpha$ is the mean time between the close particle collisions. In the hydrodynamic regime $\tau_\alpha$ can be expressed in terms of the Coulomb logarithm $\Lambda$, viscosity $\mu_\alpha$, static pressure $p_\alpha$, and a factor $\Pi$ reflecting the particle interaction model [8].

4. Many GHE applications were realized for calculation of turbulent flows with the good coincidence with the bench-mark experiments (see for example, [8]). GHE are working with good accuracy even in the theory of sound propagation in the rarefied gases where all moment equations based on the classical BE lead to unsatisfactory results.

5. The non-local kinetic effects listed above will always be relevant to a kinetic theory using one particle description – including, in particular, applications to liquids or plasmas, where self-consistent forces with appropriately cut-off radius of their action are introduced to expand the capability of GBE. The application of the above principles also leads to the modification of the system of the Maxwell electro-dynamic equations (ME). While the traditional formulation of this system does not involve the continuity equation (like (1.7) but for the charge density $\rho^a$ and the current density $\mathbf{j}^a$), nevertheless the ME derivation employs continuity equation and leads to appearance of fluctuations (proportional to $\tau$) of charge density and the current density. In rarefied media both effects lead to Johnson's flicker noise observed in 1925 for the first time by J.B. Johnson by the measurement of current fluctuations of thermo-electron emission.

Finally we can state that introduction of control volume by the reduced description for ensemble of particles of finite diameters leads to fluctuations (proportional to Knudsen number) of velocity moments in the volume. This fact leads to the significant reconstruction of the theory of transport processes. Obviously the mentioned non-local effects can be discussed from positions of breaking of the Bell's inequalities [11] because in the non-local theory the measurement (realized in $\mathbf{PhSV_1}$) influents on the measurement realized in the adjoining space-time point in $\mathbf{PhSV_2}$ and verse versa.

The generalized Boltzmann equation in general and that for plasma in particular have a fundamentally important feature that the additional GBE terms prove to be of the order of the Knudsen number. This does not mean that in the hydrodynamic (small *Kn*) limit these terms may be neglected: the Knudsen number in this case appears as a small parameter of the higher derivative in the GBE. Consequently, the additional GBE terms (as compared to the BE) are significant for any *Kn*, and the order of magnitude of the difference between the BE and GBE solutions is impossible to tell beforehand.

In this connection, it is of interest to apply the GBE model to obtain the dispersion relation for plasma in the absence of a magnetic field. This program was realized in the following sections of the article including the deep analogy existing between plasma and gravitation phenomena.

**2. Dispersion equations of plasma in the generalized Boltzmann theory**.

Let us derive the dispersion equations of plasma in the generalized Boltzmann theory using the assumptions that were used by Landau in the BE-based derivation, namely:
(a) the evolution of electrons and ions in a self-consistent electric field corresponds to a nonstationary one-dimensional model;
(b) the distribution functions for ions, $f_i$, and for electrons, $f_e$, deviate little from their equilibrium counterparts $f_{0i}$ and $f_{0e}$;

$$f_e = f_{0e}(u) + \delta f_e(x,u,t), \qquad (2.1)$$
$$f_i = f_{0i}(u) + \delta f_i(x,u,t); \qquad (2.2)$$

(c) a wave number $k$ and a complex frequency $\omega$ are appropriate to the wave mode considered; for example

$$\delta f_e = \langle \delta f_e \rangle e^{i(kx-\omega t)}, \qquad (2.3)$$

(d) the quadratic GBE terms determining the deviation from the equilibrium DF are neglected;
(e) the self-consistent forces $F_i$ and $F_e$ are sufficiently small. It particular for electron species the self-consistent force $F_e$ is not too large

$$F_e = \frac{e}{m_e} \frac{\partial \varphi}{\partial x}, \qquad (2.4)$$

where $e$ - absolute electron charge. It means that the equilibrium DF is sufficient for calculation of function $F_e\left(\partial f_e / \partial u\right)$:

$$F_e \frac{\partial f_e}{\partial u} = \frac{e}{m_e} \frac{\partial \varphi}{\partial x} \frac{\partial f_{0e}}{\partial u} \qquad (2.5)$$

(f) the change of the electrical potential corresponds to the same spatial-time dependence as the perturbation of DF

$$\varphi = \langle \varphi \rangle e^{i(kx-\omega t)}. \qquad (2.6)$$

Considering the more general case we take into account the collisional term written in the Bhatnagar - Gross - Krook (BGK) form

$$J_\alpha = -\frac{f_\alpha - f_{0\alpha}}{v_{r\alpha}^{-1}} \qquad (2.7)$$

in the right-hand side of the GBE. Here, $f_{0\alpha}$ and $v_\alpha^{-1} = \tau_{r\alpha}$ are respectively a certain equilibrium distribution function and the relaxation time for species of the $\alpha$ th kind $(\alpha = e, i)$.

Strengths and weaknesses of the BGK method are known. BGK – approximation for the local collision integral conserves (for point-like particles) the number of particles in physical system, leads to H – theorem, but violates the laws of impulse and energy conservation. The way for overcoming of these shortages is well-known and connected with application of S - method [12]. But accuracy of BGK approximation is sufficient for following investigation of asymptotic solutions. The quadratic terms determining the deviation from the equilibrium DF in kinetic equations (see (1.10))

$$\frac{\partial f_\alpha}{\partial t} + u\frac{\partial f_\alpha}{\partial x} + F_\alpha \frac{\partial f_\alpha}{\partial u} - \tau_\alpha \left\{ \frac{\partial^2 f_\alpha}{\partial t^2} + 2u\frac{\partial^2 f_\alpha}{\partial t \partial x} + u^2 \frac{\partial^2 f_\alpha}{\partial x^2} + 2F_\alpha \frac{\partial^2 f_\alpha}{\partial t \partial u} + \right.$$
$$\left. \frac{\partial F_\alpha}{\partial t}\frac{\partial f_\alpha}{\partial u} + F_\alpha \frac{\partial f_\alpha}{\partial x} + u\frac{\partial F_\alpha}{\partial x}\frac{\partial f_\alpha}{\partial u} + F_\alpha^2 \frac{\partial^2 f_\alpha}{\partial u^2} + 2uF_\alpha \frac{\partial^2 f_\alpha}{\partial u \partial x} \right\} = -v_{r\alpha} \delta f_\alpha \quad (2.8)$$

are neglected. Let us write down the complex frequency $\omega$ in the form $\omega = \omega' + i\omega''$. It means that

$$\delta f_e = \langle \delta f_e \rangle e^{i(kx - \omega' t)} e^{\omega'' t}. \quad (2.9)$$

Using assumptions formulated above, one obtains from Eq. (2.8) for electron species

$$i(ku - \omega)\langle \delta f_e \rangle + i\frac{e}{m_e} k\langle \varphi \rangle \frac{\partial f_{e0}}{\partial u} - (ku - \omega)\tau_e \left\{ -(ku - \omega)\langle \delta f_e \rangle - \langle \varphi \rangle \frac{ek}{m_e}\frac{\partial f_{e0}}{\partial u} \right\} = -v_{re}\delta f_e \quad (2.10)$$

or

$$\langle \delta f_e \rangle = -\frac{ek\langle \varphi \rangle}{m_e}\frac{\partial f_{e0}}{\partial u}\frac{i + (ku - \omega)\tau_e}{i(ku - \omega) + (ku - \omega)^2 \tau_e + v_{re}}. \quad (2.11)$$

The next step consists in application of Poisson equation for exclusion of $\langle \varphi \rangle$ from Eq. (2.11) and analogical equation for $i$ - species. Poisson equation will be used in the form

$$\Delta \varphi = -\frac{1}{\varepsilon_0}\rho, \quad (2.12)$$

then

$$\varepsilon_0 k^2 \varphi = e(\delta n_i - \delta n_e), \quad (2.13)$$

After integration over velocity $u$ we find the dispersion equation

$$1 = -\frac{e^2}{\varepsilon_0 k}\left\{ \frac{1}{m_e}\int_{-\infty}^{+\infty}\frac{(\partial f_{e0}/\partial u)[i - \tau_e(\omega - ku)]}{i(\omega - ku) - \tau_e(\omega - ku)^2 - v_{re}}du + \frac{1}{m_i}\int_{-\infty}^{+\infty}\frac{(\partial f_{i0}/\partial u)[i - \tau_i(\omega - ku)]}{i(\omega - ku) - \tau_i(\omega - ku)^2 - v_{ri}}du \right\},$$
$$(2.14)$$

originated by $e, i$ - concentrations

$$\langle \delta n_e \rangle = -\frac{ek\langle \varphi \rangle}{m_e}\int_{-\infty}^{\infty}\frac{\partial f_{e0}}{\partial u}\frac{i + (ku - \omega)\tau_e}{i(ku - \omega) + (ku - \omega)^2 \tau_e + v_{re}}du, \quad (2.15)$$

$$\langle \delta n_i \rangle = \frac{ek\langle \varphi \rangle}{m_i}\int_{-\infty}^{\infty}\frac{\partial f_{i0}}{\partial u}\frac{i + (ku - \omega)\tau_i}{i(ku - \omega) + (ku - \omega)^2 \tau_i + v_{ri}}du. \quad (2.16)$$

Interesting to notice, that for the non-collisional case $(v_{r\alpha} = 0)$ Eq. (2.14) as well as corresponding equation for $i$ species leads to the well-known relations

$$\langle \delta n_i \rangle = -\frac{e}{m_i} \langle \varphi \rangle k \int \frac{\partial f_{i0}/\partial u}{\omega - ku} du, \qquad (2.17)$$

$$\langle \delta n_e \rangle = \frac{e}{m_e} \langle \varphi \rangle k \int \frac{\partial f_{e0}/\partial u}{\omega - ku} du \qquad (2.18)$$

for the non-collisional Landau theory for all non-local parameters $\tau_e, \tau_i$. It is no surprise because the expression

$$\frac{Df_\alpha}{Dt} = 0, \ \alpha = e, i \qquad (2.19)$$

satisfies identically the equation

$$\frac{Df_\alpha}{Dt} - \frac{D}{Dt}\left(\tau_\alpha \frac{Df_\alpha}{Dt}\right) = 0. \qquad (2.20)$$

Obviously analog of Eq. (2.14) in the Boltzmann kinetic theory looks like

$$1 = -\frac{e^2}{\varepsilon_0 k}\left\{\frac{1}{m_e}\int_{-\infty}^{+\infty}\frac{(\partial f_{0e}/\partial u)}{\omega - ku + i\nu_{re}}du + \frac{1}{m_i}\int_{-\infty}^{+\infty}\frac{(\partial f_{0i}/\partial u)}{\omega - ku + i\nu_{ri}}du\right\}. \qquad (2.21)$$

Now we introduce the next Landau assumption: ions are in rest, the distribution functions (DF) for electrons $f_e$ corresponds to the maxwellian function $f_{0e}$ written for one dimensional case.

As result from Eq. (2.21)

$$1 - \frac{1}{r_D^2 k^2}\left(\frac{m_e}{2\pi k_B T}\right)^{1/2}\int_{-\infty}^{+\infty}\frac{[i - \tau(\omega - ku)] k u e^{-m_e u^2/2k_B T}}{i(\omega - ku) - \tau(\omega - ku)^2 - \nu_{re}}du = 0,$$

or after transformations

$$1 + \frac{1}{r_D^2 k^2}\left[1 - \sqrt{\frac{m_e}{2\pi k_B T}}\int_{-\infty}^{+\infty}\frac{\{[i - \tau(\omega - ku)]\omega - \nu_{re}\}e^{-m_e u^2/2k_B T}}{i(\omega - ku) - \tau(\omega - ku)^2 - \nu_{re}}du\right] = 0, \qquad (2.22)$$

where $r_D = \sqrt{\varepsilon_0 k_B T/n_e e^2}$ is Debye - Hückel radius, $k_B$ - Boltzmann constant. Eq. (2.22) is convenient to write in the dimensionless form

$$1 + \frac{1}{r_D^2 k^2}\left\{\begin{array}{l}1 + \frac{1}{\sqrt{\pi}}\left[\frac{i\tilde{\nu}_{re} + 0.5\widehat{\omega}}{\sqrt{1 + 4\widetilde{\tau}\widetilde{\nu}_{re}}} - 0.5\widehat{\omega}\right]\int_{-\infty}^{+\infty}\frac{e^{-\widehat{u}^2}}{\widehat{u}_1 - \widehat{u}}d\widehat{u} - \\ \frac{1}{\sqrt{\pi}}\left[\frac{i\tilde{\nu}_{re} + 0.5\widehat{\omega}}{\sqrt{1 + 4\widetilde{\tau}\widetilde{\nu}_{re}}} + 0.5\widehat{\omega}\right]\int_{-\infty}^{+\infty}\frac{e^{-\widehat{u}^2}}{\widehat{u}_2 - \widehat{u}}d\widehat{u}\end{array}\right\} = 0, \qquad (2.23)$$

where

$$\hat{u} = u\sqrt{\frac{m_e}{2k_BT}}, \quad \hat{\omega} = \omega \frac{1}{k}\sqrt{\frac{m_e}{2k_BT}}, \quad \hat{v}_{re} = v_{re}\frac{1}{k}\sqrt{\frac{m_e}{2k_BT}}, \quad \hat{\tau} = \tau k\sqrt{\frac{2k_BT}{m_e}} \qquad (2.24)$$

$$\hat{u}_1 = \hat{\omega} - \frac{i}{2\hat{\tau}}\left(1 + \sqrt{1 + 4\hat{\tau}\hat{v}_{re}}\right), \qquad (2.25)$$

$$\hat{u}_2 = \hat{\omega} - \frac{i}{2\hat{\tau}}\left(1 - \sqrt{1 + 4\hat{\tau}\hat{v}_{re}}\right). \qquad (2.26)$$

Eq. (2.23) can be transformed for collisionless case ($\hat{v}_{re} = 0$) in the Landau dispersion equation

$$\hat{\omega}\int_{-\infty}^{+\infty}\frac{e^{-\hat{u}^2}}{(\hat{u}_2 - \hat{u})}d\hat{u} = \sqrt{\pi}\left(1 + r_D^2 k^2\right). \qquad (2.27)$$

Equation (2.23), (2.27) contain improper Cauchy type integrals. From the theory of complex variables is known Cauchy's integral formula: if the function $f(z)$ is analytic inside and on a simple closed curve $C$, and $z_0$ is any point inside C, then

$$f(z_0) = -\frac{1}{2\pi i}\oint_C \frac{f(z)}{z_0 - z}dz \qquad (2.28)$$

where $C$ is traversed in the positive (counterclockwise) sense.

Let $C$ be the boundary of a simple closed curve placed in lower half plane (for example a semicircle of radius $R$) with the corresponding element of real axis, $z_0$ is an interior point. As usual after adding to this semicircle a cross-cut connecting semicircle $C$ with the interior circle (surrounding $z_0$) of the infinite small radius for analytic $f(z)$ the formula takes place

$$\oint_C \frac{f(z)}{z_0 - z}dz = -\int_{-R}^{R}\frac{f(\tilde{x})}{z_0 - \tilde{x}}d\tilde{x} + \int_{C_R}\frac{f(z)}{z_0 - z}dz + 2\pi i f(z_0), \qquad (2.29)$$

because the integrals along cross-cut cancel each other, ($z = \tilde{x} + i\tilde{y}$).

Analogically for upper half plane

$$\oint_C \frac{f(z)}{z_0 - z}dz = \int_{-R}^{R}\frac{f(\tilde{x})}{z_0 - \tilde{x}}d\tilde{x} + \int_{C_R}\frac{f(z)}{z_0 - z}dz + 2\pi i f(z_0). \qquad (2.30)$$

The formulae (2.29), (2.30) could be used for calculation (including the case $R \to \infty$) of the integrals along the real axis with the help of the residual theory *for arbitrary* $z_0$ if analytical function $f(z)$ satisfies the special conditions of decreasing by $R \to \infty$.

Let us consider now integral $\int_{C_R}\frac{e^{-z^2}}{z_0 - z}dz$. Generally speaking for function $f(z) = e^{-z^2}$ Cauchy's conditions are not satisfied. Really for a point $z = \tilde{x} + i\tilde{y}$ this function is $f(z) = e^{\tilde{y}^2 - \tilde{x}^2}[\cos(2\tilde{x}\tilde{y}) - i\sin(2\tilde{x}\tilde{y})]$ and by $|\tilde{y}| > |\tilde{x}|$ the function is growing for this part of $C_R$.

*But from physical point of view in* **the linear problem** *of interaction of individual electrons* **only** *with waves of potential electric field the natural assumption can be introduced that solution depends* **only** *of concrete $z_0 = \hat{\omega}' + i\hat{\omega}''$, but does not depend of another possible modes of oscillations in physical system.*

It can be realized only if the calculations do not depend of choosing of contour $C_R$. This fact leads to the additional conditions, for lower half plane

$$\int_{-\infty}^{\infty} \frac{f(\tilde{x})}{z_0 - \tilde{x}} d\tilde{x} = 2\pi i f(z_0), \tag{2.31}$$

and for upper half plane

$$\int_{-\infty}^{\infty} \frac{f(\tilde{x})}{z_0 - \tilde{x}} d\tilde{x} = -2\pi i f(z_0). \tag{2.32}$$

As it shown in [5] Landau approximation for improper integral in Eq. (2.27) contains in implicit form restrictions (valid only for close vicinity of $\tilde{x}$-axis) for the contour $C$ choosing which leads to the continuous spectrum.

The question arises, is it possible to find solutions of the equation (2.23) by the restrictions (2.31), (2.32)? In the following will be shown that the conditions (2.31), (2.32) together with Eq. (2.23) lead to the discrete spectrum of $z_0 = \hat{\omega}' + i\hat{\omega}''$ and from physical point of view conditions (2.31), (2.32) can be considered as condition of quantization.

Let us first analyze the conditions under which plasma waves can be damped. This requires, that [see Eq. (2.9)] the imaginary part of the complex frequency fulfill the condition

$$\hat{\omega}'' < 0. \tag{2.33}$$

If the collisional relaxation is not significant and $\hat{v}_{re} \to 0$, we have two equal integrals in relation (2.23), $\hat{u}_1 = \hat{u}_2$; coefficient in front of the integral containing $\hat{u}_1$ turns into zero by $\hat{v}_{re} = 0$.

Let us investigate the positions of the poles involved in the calculation of the integrals in Eq. (2.23) taking into account the conditions (2.31), (2.32):

A) The pole $\hat{u}_2$ lies in the lower half-plane

$$\hat{\omega}'' < \frac{1 - \sqrt{1 + 4\hat{\tau}\hat{v}_{re}}}{2\hat{\tau}}. \tag{2.34}$$

Then it can be realized if only

$$|\hat{\omega}''| > \left| \frac{1 - \sqrt{1 + 4\hat{\tau}\hat{v}_{re}}}{2\hat{\tau}} \right|, \tag{2.35}$$

or for small $\hat{v}_{re}$, if only

$$|\hat{\omega}''| > \hat{v}_{re}. \tag{2.36}$$

B) The pole $\hat{u}_2$ lies in the upper half-plane. Analogically for this case we have

$$|\hat{\omega}''| < \left| \frac{1 - \sqrt{1 + 4\hat{\tau}\hat{v}_{re}}}{2\hat{\tau}} \right| \tag{2.37}$$

or for small $\hat{v}_{re}$

$$|\hat{\omega}''| < \hat{v}_{re}. \tag{2.38}$$

Conditions (2.36) – (2.38) have physical sense, but lead to different expressions for conditions of quantization (2.31), (2.32), because for lower half-plane

$$I_2 = \int_{-\infty}^{+\infty} \frac{e^{-\hat{u}^2}}{\hat{u}_2 - \hat{u}} d\hat{u} = 2\pi i e^{-\hat{u}_2^2} \tag{2.39}$$

and for the upper half plane

$$I_2 = \int_{-\infty}^{+\infty} \frac{e^{-\hat{u}^2}}{\hat{u}_2 - \hat{u}} d\hat{u} = -2\pi i e^{-\hat{u}_2^2}. \tag{2.40}$$

If $\widehat{\omega}'' < 0$ the pole $\widehat{u}_1 = \widehat{\omega}' + i\left(\widehat{\omega}'' - \dfrac{1+\sqrt{1+4\widehat{\tau}\widehat{v}_{re}}}{2\widehat{\tau}}\right)$ is placed in the lower half-plane and

$$I_1 = \int_{-\infty}^{+\infty} \frac{e^{-\widehat{u}^2}}{\widehat{u}_1 - \widehat{u}} d\widehat{u} = 2\pi i e^{-\widehat{u}_1^2}. \tag{2.41}$$

In many cases of the practical significance this integral can be excluded from consideration. Really, let us write down the dispersion equation where the both poles are in the lower half-plane

$$\frac{-\widehat{v}_{re}+0.5i\widehat{\omega}}{\sqrt{1+4\widehat{\tau}\widehat{v}_{re}}} + 0.5i\widehat{\omega} - \left[\frac{-\widehat{v}_{re}+0.5i\widehat{\omega}}{\sqrt{1+4\widehat{\tau}\widehat{v}_{re}}} - 0.5i\widehat{\omega}\right]e^{\widehat{u}_2^2 - \widehat{u}_1^2} = \frac{1+r_D^2 k^2}{2\sqrt{\pi}} e^{\widehat{u}_2^2}, \tag{2.42}$$

where

$$\widehat{u}_2^2 - \widehat{u}_1^2 = \frac{\sqrt{1+4\widehat{\tau}\widehat{v}_{re}}}{\widehat{\tau}}\left[\frac{1}{\widehat{\tau}} + 2i\widehat{\omega}' - 2\widehat{\omega}''\right]. \tag{2.43}$$

If $\widehat{v}_{re} \ll 1$ or $v_{re} \ll \dfrac{2\pi}{\lambda}\sqrt{\dfrac{2k_B T}{m_e}}$ the term in square brackets close to zero and exponential term $\exp(\widehat{u}_2^2 - \widehat{u}_1^2) \sim 1$ for large mean time between close collisions.

Equation (2.42) is written for poles in the lower half-plane. If the pole $\widehat{u}_2$ is placed in the upper half plane (in spite of $\widehat{\omega}'' < 0$) the sign before the exponential term should be changed and we have unified formula in the considering case of small $\widehat{v}_{re}$.

$$\mp e^{\widehat{u}_2^2} \frac{1+r_D^2 k^2}{2\sqrt{\pi}} = \frac{\widehat{v}_{re}}{\sqrt{1+4\widehat{\tau}\widehat{v}_{re}}} - \frac{i\widehat{\omega}}{2}\left(1 + \frac{1}{\sqrt{1+4\widehat{\tau}\widehat{v}_{re}}}\right). \tag{2.44}$$

where

$$\widehat{u}_2^2 = \widehat{\omega}'^2 - \widehat{\omega}''^2 - \widehat{\omega}''\frac{\sqrt{1+4\widehat{\tau}\widehat{v}_{re}}-1}{\widehat{\tau}} - \frac{1+2\widehat{\tau}\widehat{v}_{re} - \sqrt{1+4\widehat{\tau}\widehat{v}_{re}}}{2\widehat{\tau}^2} + i\left(2\widehat{\omega}'' + \frac{\sqrt{1+4\widehat{\tau}\widehat{v}_{re}}-1}{\widehat{\tau}}\right)\widehat{\omega}'. \tag{2.45}$$

As we'll see lower, equation (2.44) admits the exact solution.

The relaxation time $\tau_{rel}$ can be estimated in terms of the mean time $\tau$ between close collisions and the Coulomb logarithm [8]:

$$\tau_{rel} = \tau\Lambda^{-1} \tag{2.46}$$

We can then write

$$\tau v_{re} = \Lambda, \widehat{\tau}\widehat{v}_{re} = \Lambda. \tag{2.47}$$

Now, in Eq. (2.44) we separate the real and imaginary parts. One obtains for the real part using (2.47)

$$\mp \frac{1+r_D^2 k^2}{2\sqrt{\pi}} \exp\left\{\widehat{\omega}'^2 - \widehat{\omega}''^2 - \widehat{\omega}''\widehat{v}_{re}\frac{\sqrt{1+4\Lambda}-1}{\Lambda} - \widehat{v}_{re}^2 \frac{1+2\Lambda - \sqrt{1+4\Lambda}}{2\Lambda^2}\right\} =$$

$$= \left[\frac{\widehat{v}_{re}}{\sqrt{1+4\Lambda}} + 0.5\widehat{\omega}'' + \frac{0.5\widehat{\omega}''}{\sqrt{1+4\Lambda}}\right]\cos\left[\widehat{\omega}'\left(2\widehat{\omega}'' + \widehat{v}_{re}\frac{\sqrt{1+4\Lambda}-1}{\Lambda}\right)\right] - \tag{2.48}$$

$$0.5\widehat{\omega}'\left[1 + \frac{1}{\sqrt{1+4\Lambda}}\right]\sin\left[\widehat{\omega}'\left(2\widehat{\omega}'' + \widehat{v}_{re}\frac{\sqrt{1+4\Lambda}-1}{\Lambda}\right)\right].$$

Similarly, for the imaginary part we find

$$0.5\hat{\omega}'\left[1+\frac{1}{\sqrt{1+4\Lambda}}\right]\cos\left[\hat{\omega}'\left(2\hat{\omega}''+\hat{v}_{re}\frac{\sqrt{1+4\Lambda}-1}{\Lambda}\right)\right]+$$
$$\left[\frac{\hat{v}_{re}}{\sqrt{1+4\Lambda}}+0.5\hat{\omega}''+\frac{0.5\hat{\omega}''}{\sqrt{1+4\Lambda}}\right]\sin\left[\hat{\omega}'\left(2\hat{\omega}''+\hat{v}_{re}\frac{\sqrt{1+4\Lambda}-1}{\Lambda}\right)\right]=0\,. \tag{2.49}$$

The system of complicated transcendent equations (2.48), (2.49) can generally be solved only on a computer. If, however, the Coulomb logarithm $\Lambda$ is large enough for terms of order $\Lambda^{-1/2}$ to be negligible or for small Coulomb logarithm $\Lambda$, the exact solution can be obtained. The system of equations (2.48), (2.49) for the large Coulomb logarithm $\Lambda$ simplifies to

$$\mp\frac{1+r_D^2 k^2}{\sqrt{\pi}}e^{\hat{\omega}'^2-\hat{\omega}''^2}=\hat{\omega}''\cos(2\hat{\omega}'\hat{\omega}'')-\hat{\omega}'\sin(2\hat{\omega}'\hat{\omega}''), \tag{2.50}$$

$$\hat{\omega}'\cos(2\hat{\omega}'\hat{\omega}'')+\hat{\omega}''\sin(2\hat{\omega}'\hat{\omega}'')=0\,. \tag{2.51}$$

Let us introduce the notation

$$\alpha=2\hat{\omega}'\hat{\omega}''\,,\ \beta=1+r_D^2 k^2 \tag{2.52}$$

and note that

$$\hat{\omega}'^2=-\frac{1}{2}\alpha\tan\alpha\,,\ \hat{\omega}''^2=-\frac{1}{2}\alpha\cot\alpha\,,\ \hat{\omega}'^2-\hat{\omega}''^2=\alpha\cot 2\alpha\,. \tag{2.53}$$

From relations (2.50) – (2.53) follow

$$\mp\frac{\beta}{\sqrt{\pi}}e^{\alpha\cot 2\alpha}=\hat{\omega}''\cos\alpha-\hat{\omega}'\sin\alpha\,. \tag{2.54}$$

Taking the square of both sides of Eq. (2.54) we obtain the universal equation

$$-e^{\sigma\cot\sigma}\sin\sigma=\frac{\pi}{2\beta^2}\sigma\,, \tag{2.55}$$

where the notation is introduced $\sigma=-2\alpha=-4\hat{\omega}'\hat{\omega}''$. As it was mentioned, this equation does not depend on the sign in front of parameter $\beta$.

The exact solution of equation (2.55) can be found with the help of the $W$-function of Lambert

$$\sigma_n=\text{Im}\left[W_n\left(\frac{2\beta^2}{\pi}\right)\right], \tag{2.56}$$

frequencies $\hat{\omega}'_n,\hat{\omega}''_n$ are (see (2.53))

$$\omega'_n=k\sqrt{-\frac{k_B T}{2m_e}\sigma_n\tan\frac{\sigma_n}{2}}\,,\ \omega''_n=-k\sqrt{-\frac{k_B T}{2m_e}\sigma_n\cot\frac{\sigma_n}{2}}\,. \tag{2.57}$$

In asymptotic for large entire positive $n$

$$\sigma_n=\left(n+\frac{1}{2}\right)\pi\,,\ \hat{\omega}'_n=\frac{\sqrt{\sigma_n}}{2}=\frac{1}{2}\sqrt{\pi\left(n+\frac{1}{2}\right)}\,,\ \hat{\omega}''_n=-\frac{\sqrt{\sigma_n}}{2}=-\frac{1}{2}\sqrt{\pi\left(n+\frac{1}{2}\right)}. \tag{2.58}$$

The exact solution for the $n$ – th discrete solution from the spectrum of oscillations follows from (2.56), (2.57):

$$\hat{\omega}_n = \frac{1}{2}\sqrt{-\mathrm{Im}\left[W_n\left(\frac{2(1+r_D^2k^2)^2}{\pi}\right)\right]\tan\left[\frac{1}{2}\mathrm{Im}\left[W_n\left(\frac{2(1+r_D^2k^2)^2}{\pi}\right)\right]\right]} -$$
$$\frac{i}{2}\sqrt{-\mathrm{Im}\left[W_n\left(\frac{2(1+r_D^2k^2)^2}{\pi}\right)\right]\cot\left[\frac{1}{2}\mathrm{Im}\left[W_n\left(\frac{2(1+r_D^2k^2)^2}{\pi}\right)\right]\right]}.$$

(2.59)

The square of the oscillation frequency of plasma waves $\omega_n'^2$ is proportional to the wave energy. Hence, the energy of plasma waves is quantized, and as $n$ grows we have the asymptotic expression

$$\hat{\omega}_n'^2 = \frac{\pi}{4}\left(n+\frac{1}{2}\right),$$

(2.60)

the squares of possible dimensionless frequencies become equally spaced:

$$\hat{\omega}_{n+1}'^2 - \hat{\omega}_n'^2 = \frac{\pi}{4}.$$

(2.61)

Figures 2.1 and 2.2 reflect the result of calculations for 200 discrete levels for the case of the large Coulomb logarithm $\Lambda$.

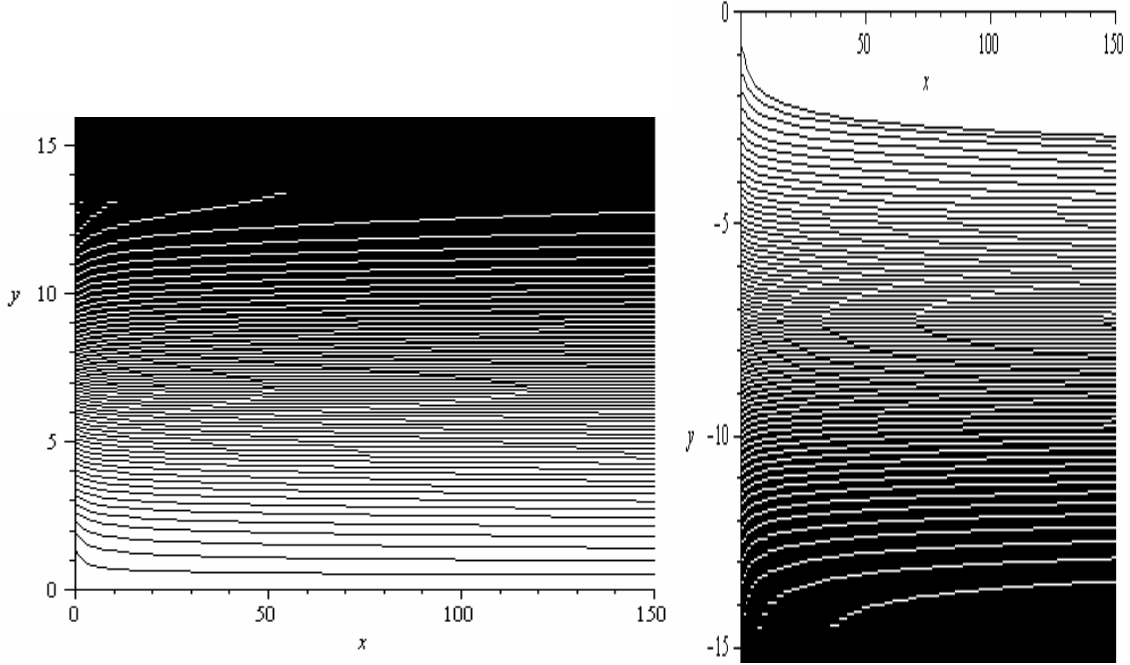

Fig. 2.1. The dimensionless frequency $\hat{\omega}'$ ($y$ axes) versus parameter $r_D k$ ($x$ axes); (left).
Fig. 2.2. The dimensionless frequency $\hat{\omega}''$ ($y$ axes) versus parameter $r_D k$ ($x$ axes); (right).

Plotter from the technical point of view has no possibility to reflect the small curvature of lines and approximates this curvature as a long step with shifting to the next pick cell. For high levels this spectrum contains many very close practically straight lines with small but long steps, which human eyes can perceive as background. As result plotters are realizing the discrete constriction of derivatives $d(r_D k)/d\hat{\omega}'$ and $d(r_D k)/d\hat{\omega}''$ for discontinuous functions. This effect has no attitude to the mathematical programming. You can see this very complicated topology of curves including the spectrum of the bell-like dispersion curves in Fig. 2.1 and Fig. 2.2, which also form the discrete spectrum.

But mentioned derivatives can be written in the form

$$\frac{d(r_D k)}{d\hat{\omega}'} = \frac{dk}{d(\omega'/k)} \frac{k_B T \sqrt{2}}{m_e \omega_e} = \frac{dk}{dv_\phi} \frac{k_B T \sqrt{2}}{m_e \omega_e}, \qquad (2.62)$$

where $v_\phi$ is the wave phase velocity. Then

$$\frac{dv_\phi}{d\lambda} = -\frac{1}{\lambda^2} \frac{k_B T}{e} \sqrt{\frac{2\pi}{\rho_e}} \left[\frac{d(r_D k)}{d\hat{\omega}'}\right]^{-1}, \qquad (2.63)$$

where $\rho_e = m_e n_e$. Very complicated topology of the dispersion curves exist also in the black domains of Fig. 2.1, Fig. 2.2 where (in chosen scale) the curves cannot be observed separately. Then Figs. 2.1 and 2.2 can be used for understanding of the future development of events in physical system after the initial linear stage. For example Fig. 2.1 shows the discrete set of frequencies which vicinity corresponds to passing over from abnormal to normal dispersion (for example, by $\hat{\omega}' \sim 7$) for discrete systems of $r_D k$. These discrete systems of $r_D k$ have regular character with factor of about two. Fig. 2.3 demonstrates the close up view of the mentioned discrete system of passing over from abnormal to normal dispersion in the range between 0 and 50 $r_D k$. Of course the non-linear stage needs the special investigation with using of other methods including the method of direct mathematical modeling.

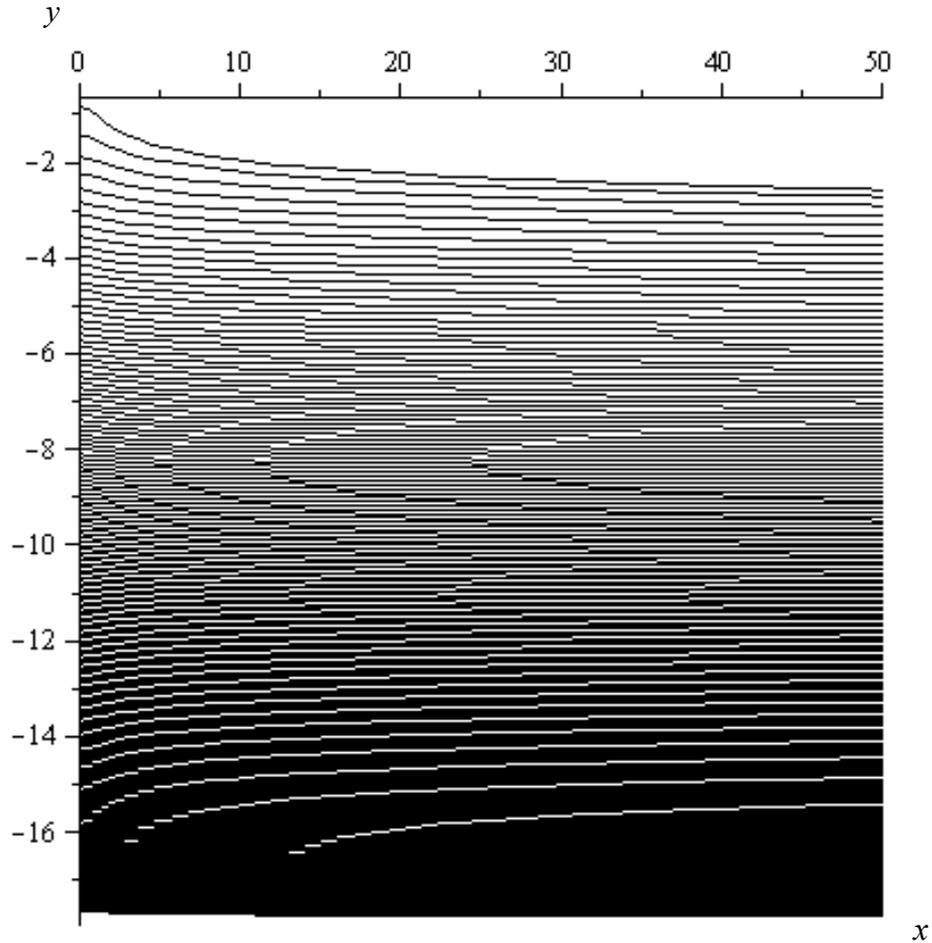

Fig. 2.3. The dimensionless frequency $\widehat{\omega}''$ ($y$ axes) versus parameter $r_D k$ ($x$ axes).

Let us investigate now the case of small Coulomb logarithm $\Lambda$ using Eqs. (2.48), (2.49). One obtains for real and imaginary parts correspondingly

$$\mp \frac{1 + r_D^2 k^2}{2\sqrt{\pi}} \exp\{\widehat{\omega}'^2 - [\widehat{\omega}'' + \widehat{v}_{re}]^2\} = \tag{2.64}$$
$$= [\widehat{\omega}'' + \widehat{v}_{re}] \cos[2\widehat{\omega}'(\widehat{\omega}'' + \widehat{v}_{re})] - \widehat{\omega}' \sin[2\widehat{\omega}'(\widehat{\omega}'' + \widehat{v}_{re})],$$

$$\widehat{\omega}' \cos[2\widehat{\omega}'(\widehat{\omega}'' + \widehat{v}_{re})] + (\widehat{\omega}'' + \widehat{v}_{re})\sin[2\widehat{\omega}'(\widehat{\omega}'' + \widehat{v}_{re})] = 0. \tag{2.65}$$

Equations (2.64) and (2.65) have practically the same form as the system of equations (2.50), (2.51) after replacing $\widehat{\omega}''$ with the variable $\widehat{\omega}_1'' = \widehat{\omega}'' + \widehat{v}_{re}$. It should also be noted that, in the asymptotical case we are considering, an additional factor 0.5 appears on the left-hand side of Eqn (2.64). Equations (2.64) and (2.65) are then solved in exactly the same manner to give (for large $n$):

$$\widehat{\omega}_n' = \frac{1}{2}\sqrt{\pi\left(n + \frac{1}{2}\right)}, \widehat{\omega}_n'' = -\frac{1}{2}\sqrt{\pi\left(n + \frac{1}{2}\right)} - \widehat{v}_{re}. \tag{2.66}$$

Then the problem reduces to that of solving the transcendent equation

$$-e^{\sigma_1 \cot \sigma_1} \sin \sigma_1 = \frac{2\pi}{\beta^2} \sigma_1, \tag{2.67}$$

where

$$\sigma_1 = -4\widehat{\omega}'\widehat{\omega}_1'', \quad \widehat{\omega}_1'' = \widehat{\omega}'' + \widehat{v}_{re}. \tag{2.68}$$

After introducing $\widehat{v}_{re} \to 0$ in Eqs. (2.67), (2.68) we realize the passage to the generalized dispersion equation considered in [5] for the non-collisional case.

The exact solution of equation (2.67) can be found with the help of the $W$-function of Lambert

$$\sigma_{1n} = \text{Im}\left[W_n\left(\frac{\beta^2}{2\pi}\right)\right], \tag{2.69}$$

frequencies $\widehat{\omega}_n', \widehat{\omega}_n''$ are

$$\omega_n' = k\sqrt{-\frac{k_B T}{2m_e}\sigma_{1n} \tan\frac{\sigma_{1n}}{2}}, \quad \omega_{1n}'' = -k\sqrt{-\frac{k_B T}{2m_e}\sigma_{1n} \cot\frac{\sigma_{1n}}{2}}. \tag{2.70}$$

Figures 2.4 and 2.5 reflect the result of calculations for 200 discrete levels for the case of the small Coulomb logarithm $\Lambda$. We see the same character features of the topology as in the case of large Coulomb logarithm $\Lambda$ reflected on figures 2.1 – 2.3, but of course with other quantitative characteristics.

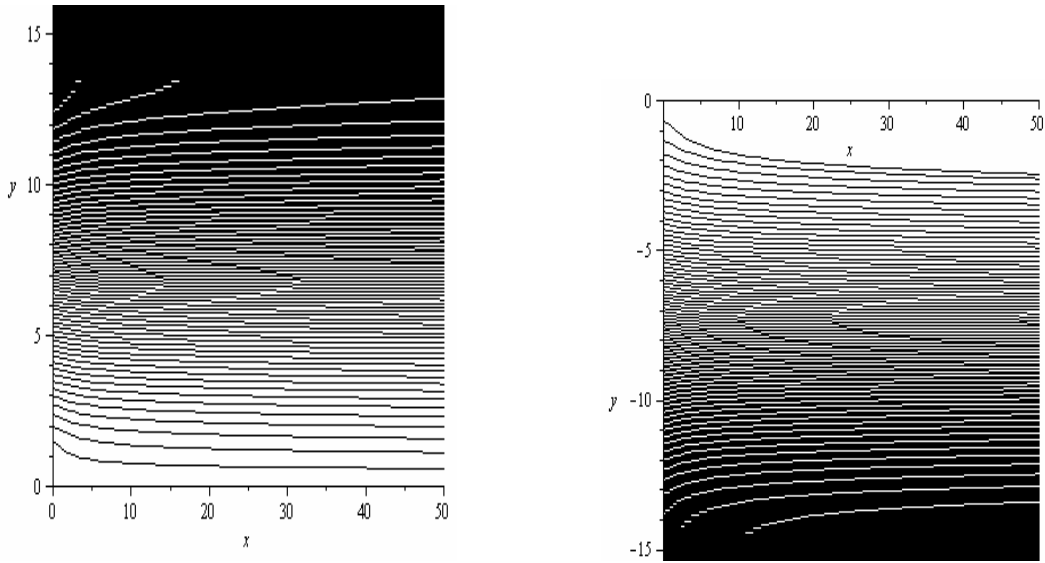

Fig. 2.4. The dimensionless frequency $\bar\omega'$ ($y$ axes) versus parameter $r_D k$ ($x$ axes); (left).
Fig. 2.5. The dimensionless frequency $\bar\omega_1''$ ($y$ axes) versus parameter $r_D k$ ($x$ axes); (right).

Dispersion equations (2.55) and (2.67) can be written in the unified form:
$$-e^{\sigma \cot \gamma} \sin \sigma = \xi \sigma, \qquad (2.71)$$

which solution is

$$\sigma_n = \mathrm{Im}\left[W_n\left(\frac{1}{\xi}\right)\right], \qquad (2.72)$$

where $\xi = \dfrac{\pi}{2\beta^2}$ for large Coulomb logarithm $\Lambda$ and $\xi = \dfrac{2\pi}{\beta^2}$ for small Coulomb logarithm $\Lambda$, $\beta = 1 + r_D^2 k^2$. Both considered cases correspond to assumption of small values $v_{re}$, in other words – to small influence of local collision integral on solution of kinetic equation. If this influence is not small, dispersion equation should contains omitted integral $I_2$ (see Eq. (2.23)) and should be written with taking into account the fluctuation effects in Poisson equation [6]. This consideration cannot be realized in the analytical form.

### 3. Plasma – gravitational analogy in the generalized theory of Landau damping.

Plasma – gravitational analogy is well-known and frequently used effect in physical kinetics. The origin of analogy is simple and is connected with analogy between Coulomb law and Newtonian law of gravitation. From other side the electrical charges can have different signs whereas there is just one kind of "gravitational charge" (i.e. masses of particles) corresponding to the force of attraction. This fact leads to the extremely important distinctions in formulation of the generalized theory of Landau damping in gravitational media.

In the following we intend to use the classical non-relativistic Newtonian law of gravitation

$$\mathbf{F}_{21} = \gamma_N \frac{m_1 m_2}{r_{12}^2} \frac{\mathbf{r}_{12}}{r_{12}}, \qquad (3.1)$$

where $\mathbf{F}_{21}$ is the force acting on the particle "1" from the particle "2", $\mathbf{r}_{12}$ is vector directed from the center-of-mass of the particle "1" to the particle "2", $\gamma_N$ is gravitational constant; the corresponding force $\mathbf{g}_{21}$ per mass unit is

$$\mathbf{g}_{21} = \frac{\mathbf{F}_{21}}{m_1}. \qquad (3.2)$$

The flux

$$\Phi = \int_S g_n dS \qquad (3.3)$$

for closed surface $S$ can be calculated using (3.1); one obtains

$$\int_S g_n dS = -4\pi\gamma_N \int_V \rho^a dV, \qquad (3.4)$$

where $\rho^a$ is density *inside* of volume $V$ bounded by the surface $S$. As usual Eq. (3.4) can be rewritten as

$$\int_V \left( \text{div } \mathbf{g} + 4\pi\gamma_N \rho^a \right) dV = 0. \qquad (3.5)$$

The definite integral (3.5) is equal to zero for arbitrary volume $V$, then

$$\text{div } \mathbf{g} = -4\pi\gamma_N \rho^a, \qquad (3.6)$$

and after introduction the gravitational potential $\Psi$

$$\mathbf{g} = -\frac{\partial \Psi}{\partial \mathbf{r}} \qquad (3.7)$$

we reach the known Poisson equation

$$\Delta \Psi = 4\pi\gamma_N \rho^a. \qquad (3.8)$$

Generalized Boltzmann physical kinetics leads to possibility to calculate the density $\rho^a$ using the density $\rho$ (obtained with the help of the one particle distribution function $f$) and the fluctuation term $\rho^{fl}$. All fluctuation terms in the GBE theory were tabulated [8] and for $\rho^{fl}$ we have

$$\rho^{fl} = \tau\left( \frac{\partial \rho}{\partial t} + \frac{\partial}{\partial \mathbf{r}} \cdot \rho\mathbf{v}_0 \right), \qquad (3.9)$$

where $\mathbf{v}_0$ is hydrodynamic velocity. After substitution of $\rho^{fl}$ in (3.8) one obtains

$$\Delta \Psi = 4\pi\gamma_N \left[ \rho - \tau\left( \frac{\partial \rho}{\partial t} + \frac{\partial}{\partial \mathbf{r}} \cdot \rho\mathbf{v}_0 \right) \right]. \qquad (3.10)$$

From Eqns (3.8) - (3.10) follows that classical Newtonian field equation

$$\Delta \Psi = 4\pi\gamma_N \rho \qquad (3.11)$$

valid only for situation when the fluctuations terms can be omitted and then

$$\rho = \rho^a. \qquad (3.12)$$

This condition can be considered as the simplest closure condition but in the general case the other hydrodynamic equations should be involved into consideration because Eq. (3.10) contains hydrodynamic velocity $\mathbf{v}_0$. As result one obtains the system of moment equations, which is written here for simplicity for maxwellian distribution function. In the general case the generalized Enskog equations should be used [8].

(gravitation equation)

$$\frac{\partial}{\partial \mathbf{r}} \cdot \mathbf{g} = -4\pi\gamma_N \left[ \rho - \tau\left( \frac{\partial \rho}{\partial t} + \frac{\partial}{\partial \mathbf{r}} \cdot \rho\mathbf{v}_0 \right) \right], \qquad (3.13)$$

(continuity equation)

$$\frac{\partial}{\partial t}\left\{\rho - \Pi\frac{\mu}{p}\left[\frac{\partial\rho}{\partial t} + \frac{\partial}{\partial \mathbf{r}}\cdot(\rho\mathbf{v}_0)\right]\right\} + \frac{\partial}{\partial \mathbf{r}}\cdot\left\{\rho\mathbf{v}_0 - \Pi\frac{\mu}{p}\left[\frac{\partial}{\partial t}(\rho\mathbf{v}_0) + \frac{\partial}{\partial \mathbf{r}}\cdot\rho\mathbf{v}_0\mathbf{v}_0 + \vec{\mathbf{I}}\cdot\frac{\partial p}{\partial \mathbf{r}} - \rho\mathbf{g}\right]\right\} = 0, \quad (3.14)$$

(motion equation)

$$\frac{\partial}{\partial t}\left\{\rho\mathbf{v}_0 - \Pi\frac{\mu}{p}\left[\frac{\partial}{\partial t}(\rho\mathbf{v}_0) + \frac{\partial}{\partial \mathbf{r}}\cdot\rho\mathbf{v}_0\mathbf{v}_0 + \frac{\partial p}{\partial \mathbf{r}} - \rho\mathbf{g}\right]\right\} - \mathbf{g}\left[\rho - \Pi\frac{\mu}{p}\left(\frac{\partial\rho}{\partial t} + \frac{\partial}{\partial \mathbf{r}}\cdot(\rho\mathbf{v}_0)\right)\right] + \frac{\partial}{\partial \mathbf{r}}\cdot\left\{\rho\mathbf{v}_0\mathbf{v}_0 + p\vec{\mathbf{I}} - \Pi\frac{\mu}{p}\left[\frac{\partial}{\partial t}(\rho\mathbf{v}_0\mathbf{v}_0 + p\vec{\mathbf{I}}) + \frac{\partial}{\partial \mathbf{r}}\cdot\rho(\mathbf{v}_0\mathbf{v}_0)\mathbf{v}_0 + 2\frac{\partial}{\partial \mathbf{r}}\left[\frac{\partial}{\partial \mathbf{r}}\cdot(p\mathbf{v}_0)\right] + \Delta(p\mathbf{v}_0) - \mathbf{g}\rho\mathbf{v}_0 - \mathbf{v}_0\mathbf{g}\rho\right]\right\} = 0, \quad (3.15)$$

(energy equation)

$$\frac{\partial}{\partial t}\left\{\frac{\rho v_0^2}{2} + \frac{3}{2}p - \Pi\frac{\mu}{p}\left[\frac{\partial}{\partial t}\left(\frac{\rho v_0^2}{2} + \frac{3}{2}p\right) + \frac{\partial}{\partial \mathbf{r}}\cdot\left(\frac{1}{2}\rho v_0^2\mathbf{v}_0 + \frac{5}{2}p\mathbf{v}_0\right) - \mathbf{g}\cdot\rho\mathbf{v}_0\right]\right\} + \frac{\partial}{\partial \mathbf{r}}\cdot\left\{\frac{1}{2}\rho v_0^2\mathbf{v}_0 + \frac{5}{2}p\mathbf{v}_0 - \Pi\frac{\mu}{p}\left[\frac{\partial}{\partial t}\left(\frac{1}{2}\rho v_0^2\mathbf{v}_0 + \frac{5}{2}p\mathbf{v}_0\right) + \frac{\partial}{\partial \mathbf{r}}\cdot\left(\frac{1}{2}\rho v_0^2\mathbf{v}_0\mathbf{v}_0 + \frac{7}{2}p\mathbf{v}_0\mathbf{v}_0 + \frac{1}{2}pv_0^2\vec{\mathbf{I}} + \frac{5}{2}\frac{p^2}{\rho}\vec{\mathbf{I}}\right) - \rho\mathbf{g}\cdot\mathbf{v}_0\mathbf{v}_0 - p\mathbf{g}\cdot\vec{\mathbf{I}} - \frac{1}{2}\rho v_0^2\mathbf{g} - \frac{3}{2}\mathbf{g}p\right]\right\} - \left\{\rho\mathbf{g}\cdot\mathbf{v}_0 - \Pi\frac{\mu}{p}\left[\mathbf{g}\cdot\left(\frac{\partial}{\partial t}(\rho\mathbf{v}_0) + \frac{\partial}{\partial \mathbf{r}}\cdot\rho\mathbf{v}_0\mathbf{v}_0 + \frac{\partial}{\partial \mathbf{r}}\cdot p\vec{\mathbf{I}} - \rho\mathbf{g}\right)\right]\right\} = 0, \quad (3.16)$$

where in hydrodynamic approximation for the maxwellian distribution function $\tau^{(0)} = \Pi\frac{\mu}{p}$ (for the hard spheres model $\Pi = 0.8$), $\mu$ is dynamic viscosity, $p$ is static pressure, $\vec{\mathbf{I}}$ is unit tensor.

As we see from the system of equations (3.14) – (3.16) gravitational force **g** (or gravitational potential $\Psi$) can be calculated only as result of the self-consistent solution of the mentioned system of moment equations.

But here I don't intend to investigate applications of the generalized hydrodynamic equations. As it will be shown the significant results can be obtained on the level of the generalized theory of Landau damping based on the generalized Boltzmann equation (GBE). The following mathematical transformations need in some preliminary additional explanations from positions of so called dark energy and dark matter.

About ten years ago the accelerated cosmological expansion was discovered in direct astronomical observations at distances of a few billion light years, almost at the edge of the observable Universe. This acceleration should be explained because mutual attraction of cosmic bodies is only capable of decelerating their scattering. As result new idea was introduced in physics about existing of a force with the opposite sign which is called universal antigravitation. Its physical source is so called dark energy that manifest itself only because of postulated property of providing antigravitation.

In simplest interpretation dark energy is related usually to the Einstein cosmological constant. In review [13] the modified Newton force is written as

$$F(r) = -\frac{\gamma_N M}{r^2} + \frac{8\pi\gamma_N}{3}\rho_{\mathbf{v}} r, \qquad (3.17)$$

where $\rho_{\mathbf{v}}$ is the Einstein – Gliner vacuum density introduced also in [13, 14]. In the limit of large distances the influence of central mass $M$ becomes negligibly small and the field of forces is determined only by the second term in the right side of (3.17). It follows from relation (3.17) that there is a distance $r_{\mathbf{v}}$ at which the sum of the gravitation and antigravitation forces is equal to zero. In other words $r_{\mathbf{v}}$ is "the zero-gravitational radius". For so called Local Group of galaxies estimation of $r_{\mathbf{v}}$ is about 1Mpc, [13]. Obviously the second term in relation (3.17) should be defined as result of solution of the self-consistent gravitational problem.

Let us return now to the formulation of plasma-gravitational analogy in the frame of generalized theory of Landau damping. I intend to apply the GBE model with the aim to obtain the dispersion relation for one component gas placed in the self-consistent gravitational field and to consider the effect of antigravitation in the frame of the Newton theory of gravitation.

With this aim let us admit now that there is a gravitational perturbation $\delta\Psi$ in the system of particles connected with the density perturbation $\delta\rho$. These perturbations connected with the perturbation of DF in the system, which was before in local equilibrium.

In doing so, we will make the additional assumptions for simplification of the problem, namely:

(a) Consideration of the self-consistent gravitational field correspond to the area of the large distance $r$ (see relation (3.17)) from the central mass $M$ where the first term is not significant and in particular the problem correspond to the plane case. As mentioned above the second term should be defined as a self-consistent force

$$F = -\frac{\partial\delta\Psi}{\partial x}. \qquad (3.18)$$

(b) The integral collision term is written in the Bhatnagar - Gross - Krook (BGK) form

$$J = -\frac{f - f_0}{v_r^{-1}} \qquad (3.19)$$

into the right-hand side of the GBE. Here, $f_0$ and $v_r^{-1} = \tau_r$ are respectively a certain equilibrium distribution function and the relaxation time.

(c) The evolution of particles in a self-consistent gravitational field corresponds to a non-stationary one-dimensional model;

(d) The distribution function $f$ deviate little from its equilibrium counterpart $f_0$.

$$f = f_0(u) + \delta f(x,u,t), \qquad (3.20)$$

(e) A wave number $k$ and a complex frequency $\omega$ ($\omega = \omega' + i\omega''$) are appropriate to the wave mode considered;

$$\delta f = \langle \delta f \rangle e^{i(kx-\omega t)}, \quad (3.21)$$

$$\delta \Psi = \langle \delta \varphi \rangle e^{i(kx-\omega t)}; \quad (3.22)$$

(f) the quadratic GBE terms determining the deviation from the equilibrium DF are neglected.

Under these assumptions listed above, the GBE is written as follows (see also (2.8)):

$$\frac{\partial f}{\partial t} + u\frac{\partial f}{\partial x} + F\frac{\partial f}{\partial u} - \tau\left\{\frac{\partial^2 f}{\partial t^2} + 2u\frac{\partial^2 f}{\partial t \partial x} + u^2\frac{\partial^2 f}{\partial x^2} + 2F\frac{\partial^2 f}{\partial t \partial u} + \frac{\partial F}{\partial t}\frac{\partial f}{\partial u} + F\frac{\partial f}{\partial x} + u\frac{\partial F}{\partial x}\frac{\partial f}{\partial u} + F^2\frac{\partial^2 f}{\partial u^2} + 2uF\frac{\partial^2 f}{\partial u \partial x}\right\} = -v_r \delta f, \quad (3.23)$$

where the relations take place

$$\frac{\partial f}{\partial t} = -i\omega \delta f, \quad u\frac{\partial f}{\partial x} = iku\delta f, \quad F\frac{\partial f}{\partial u} = -\frac{\partial \delta \Psi}{\partial x}\frac{\partial f_0}{\partial u},$$

$$\frac{\partial^2 f}{\partial t^2} = -\omega^2 \delta f, \quad 2u\frac{\partial^2 f}{\partial t \partial x} = 2\omega u k \delta f, \quad u^2 \frac{\partial^2 f}{\partial x^2} = -u^2 k^2 \delta f, \quad 2F\frac{\partial^2 f}{\partial t \partial u} = 0, \quad (3.24)$$

$$\frac{\partial F}{\partial t}\frac{\partial f}{\partial u} = -\frac{\partial}{\partial t}\frac{\partial \delta \Psi}{\partial x}\frac{\partial f}{\partial u} = -\omega k \delta \Psi \frac{\partial f_0}{\partial u}, \quad F\frac{\partial f}{\partial x} = 0, \quad u\frac{\partial f}{\partial u}\frac{\partial F}{\partial x} = -u\frac{\partial f}{\partial u}\frac{\partial}{\partial x}\frac{\partial \delta \Psi}{\partial x} = k^2 u \delta \Psi \frac{\partial f_0}{\partial u}$$

$$F^2 \frac{\partial^2 f}{\partial u^2} = 0, \quad \frac{\partial^2 f}{\partial u \partial x} 2uF = 0.$$

We are concerned with developing (within the GBE framework) the dispersion relation for gravitational field and substitution of (3.24) into Eq. (3.23) yields

$$\{i(ku-\omega) + v_r + \tau(ku-\omega)^2\}\langle \delta f \rangle - \langle \delta \Psi \rangle \frac{\partial f_0}{\partial u} k\{i + \tau(ku-\omega)\} = 0. \quad (3.25)$$

For the physical system under consideration the influence of collision term $v_r \langle \delta f \rangle$ is rather small. Using for this case the Poisson equation in the form (compare with Eq. (2.12))

$$\Delta \Psi = 4\pi \gamma_N \rho \quad (3.26)$$

and then the relation

$$k^2 \langle \delta \Psi \rangle = -4\pi \gamma_N \langle \delta n \rangle, \quad (3.27)$$

one obtains from (3.25), (3.27)

$$\langle \delta f \rangle = \frac{4\pi \gamma_N m}{k} \frac{[i - \tau(\omega - ku)]\frac{\partial f_0}{\partial u}}{i(\omega - ku) - \tau(\omega - ku)^2 - v_r} \langle \delta n \rangle. \quad (3.28)$$

After integration over all $u$ we arrive at the dispersion relation

$$1 = \frac{4\pi \gamma_N m}{k} \int_{-\infty}^{+\infty} \frac{\frac{\partial f_0}{\partial u}[i - \tau(\omega - ku)]}{i(\omega - ku) - \tau(\omega - ku)^2 - v_r} du. \quad (3.29)$$

Let us suppose that the velocity depending part of DF $f_0$ corresponds to the Maxwell DF. Then after differentiating in Eq. (3.29) under the sign of integral and some transformations we obtain the integral dispersion equation

$$1 - \frac{1}{r_a^2 k^2}\left[1 - \sqrt{\frac{m}{2\pi k_B T}} \int_{-\infty}^{+\infty} \frac{\{[i - \tau(\omega - ku)]\omega - v_r\}e^{-mu^2/2k_B T}}{i(\omega - ku) - \tau(\omega - ku)^2 - v_r} du\right] = 0, \quad (3.30)$$

where

$$r_a = \sqrt{\frac{k_B T}{4\pi \gamma_N m^2 n}}. \quad (3.31)$$

Poisson equation (3.26) has the structure like the Poisson equation for the electrical potential, as result the relation for $r_a$ is analogous to the Debye - Hueckel radius $r_D = \sqrt{k_B T/(4\pi e^2 n)}$.

Introducing now the dimensionless variables

$$\hat{u} = u\sqrt{\frac{m}{2k_B T}}, \quad \hat{\omega} = \omega \frac{1}{k}\sqrt{\frac{m}{2k_B T}}, \quad \hat{v}_r = v_r \frac{1}{k}\sqrt{\frac{m}{2k_B T}}, \quad \hat{\tau} = \tau k \sqrt{\frac{2k_B T}{m}} \quad (3.32)$$

we can rewrite Eq (3.30) in the form

$$1 - \frac{1}{r_a^2 k^2}\left[1 - \frac{1}{\sqrt{\pi}}\int_{-\infty}^{+\infty}\frac{\{[i - \hat{\tau}(\hat{\omega} - \hat{u})]\hat{\omega} - \hat{v}_r\}e^{-\hat{u}^2}}{i(\hat{\omega} - \hat{u}) - \hat{\tau}(\hat{\omega} - \hat{u})^2 - \hat{v}_r}d\hat{u}\right] = 0. \quad (3.33)$$

Now consider a situation in which the denominator of the complex integrand in Eq. (3.33) becomes zero. The quadratic equation

$$\hat{\tau} y^2 - iy + \hat{v}_r = 0, \quad y = \hat{\omega} - \hat{u} \quad (3.34)$$

has the roots

$$y_1 = \frac{i}{2\hat{\tau}}\left(1 + \sqrt{1 + 4\hat{\tau}\hat{v}_r}\right), \quad y_2 = \frac{i}{2\hat{\tau}}\left(1 - \sqrt{1 + 4\hat{\tau}\hat{v}_r}\right) \quad (3.35)$$

Hence, Eqn (3.33) can be rewritten as

$$1 - \frac{1}{r_a^2 k^2}\left[1 + \frac{1}{\hat{\tau}\sqrt{\pi}}\int_{-\infty}^{+\infty}\frac{\{[i + \hat{\tau}(\hat{u} - \hat{\omega})]\hat{\omega} - \hat{v}_r\}e^{-\hat{u}^2}}{(\hat{u} - \hat{u}_1)(\hat{u} - \hat{u}_2)}d\hat{u}\right] = 0 \quad (3.36)$$

where

$$\hat{u}_1 = \hat{\omega} - y_1, \quad \hat{u}_2 = \hat{\omega} - y_2. \quad (3.37)$$

Let us transform equation (3.33) to the following one:

$$1 - \frac{1}{r_a^2 k^2}\left\{1 + \frac{1}{\sqrt{\pi}}\left[\left(\frac{i\hat{v}_r + 0.5\hat{\omega}}{\sqrt{1 + 4\hat{\tau}\hat{v}_r}} - 0.5\hat{\omega}\right)\int_{-\infty}^{+\infty}\frac{e^{-\hat{u}^2}}{(\hat{u}_1 - \hat{u})}d\hat{u} - \right.\right.$$
$$\left.\left.\left(\frac{i\hat{v}_r + 0.5\hat{\omega}}{\sqrt{1 + 4\hat{\tau}\hat{v}_r}} + 0.5\hat{\omega}\right)\int_{-\infty}^{+\infty}\frac{e^{-\hat{u}^2}}{(\hat{u}_2 - \hat{u})}d\hat{u}\right]\right\} = 0 \quad (3.38)$$

The last equation is analog of Eq. (2.23) and contains improper Cauchy type integrals, which can be evaluated using the theory of residues. The relations (2.39) – (2.41) are the additional conditions which physical sense consists in the extraction of independent oscillations

– oscillations which existence does not depend on presence of other oscillations in considering physical system.

As in the plasma variant theory considered above, relations (2.39) – (2.41) are the conditions of quantization which consist with non-local quantum physics created in [5 - 7].

Then Eqn (3.38) produces the dispersion relation, which admits a damped gravitational wave solution $(\hat{\omega}'' < 0)$ for small influence of the collision integral (compare with (2.44)):

$$\mp e^{\hat{u}_2^2} \frac{1 - r_a^2 k^2}{2\sqrt{\pi}} = \frac{\hat{v}_r}{\sqrt{1 + 4\tilde{\tau}\hat{v}_r}} - \frac{i\hat{\omega}}{2}\left(1 + \frac{1}{\sqrt{1 + 4\tilde{\tau}\hat{v}_r}}\right). \tag{3.39}$$

where

$$\hat{u}_2^2 = \hat{\omega}'^2 - \hat{\omega}''^2 - \hat{\omega}'' \frac{\sqrt{1 + 4\tilde{\tau}\hat{v}_r} - 1}{\tilde{\tau}} - \hat{v}_r^2 \frac{1 + 2\tilde{\tau}\hat{v}_r - \sqrt{1 + 4\tilde{\tau}\hat{v}_r}}{2\tilde{\tau}^2} + i\left(2\hat{\omega}'' + \frac{\sqrt{1 + 4\tilde{\tau}\hat{v}_r} - 1}{\tilde{\tau}}\right)\hat{\omega}'. \tag{3.40}$$

As before the time of the collision relaxation $\tau_{rel} = v_r^{-1}$ for gravitational physical system can be estimated in terms of the mean time $\tau$ between close collisions and the Coulomb logarithm [8]:

$$\tau v_r = \Lambda, \tilde{\tau}\hat{v}_r = \Lambda. \tag{3.41}$$

We separate the real and imaginary parts in Eq. (3.39). One obtains for the real part

$$\mp \frac{1 - r_a^2 k^2}{2\sqrt{\pi}} \exp\left\{\hat{\omega}'^2 - \hat{\omega}''^2 - \hat{\omega}''\hat{v}_r \frac{\sqrt{1 + 4\Lambda} - 1}{\Lambda} - \hat{v}_r^2 \frac{1 + 2\Lambda - \sqrt{1 + 4\Lambda}}{2\Lambda^2}\right\} =$$

$$= \left[\frac{\hat{v}_r}{\sqrt{1 + 4\Lambda}} + 0.5\hat{\omega}'' + \frac{0.5\hat{\omega}''}{\sqrt{1 + 4\Lambda}}\right] \cos\left[\hat{\omega}'\left(2\hat{\omega}'' + \hat{v}_r \frac{\sqrt{1 + 4\Lambda} - 1}{\Lambda}\right)\right] - \tag{3.42}$$

$$0.5\hat{\omega}'\left[1 + \frac{1}{\sqrt{1 + 4\Lambda}}\right] \sin\left[\hat{\omega}'\left(2\hat{\omega}'' + \hat{v}_r \frac{\sqrt{1 + 4\Lambda} - 1}{\Lambda}\right)\right].$$

Similarly, for the imaginary part we find

$$0.5\hat{\omega}'\left[1 + \frac{1}{\sqrt{1 + 4\Lambda}}\right] \cos\left[\hat{\omega}'\left(2\hat{\omega}'' + \hat{v}_r \frac{\sqrt{1 + 4\Lambda} - 1}{\Lambda}\right)\right] +$$

$$\left[\frac{\hat{v}_r}{\sqrt{1 + 4\Lambda}} + 0.5\hat{\omega}'' + \frac{0.5\hat{\omega}''}{\sqrt{1 + 4\Lambda}}\right] \sin\left[\hat{\omega}'\left(2\hat{\omega}'' + \hat{v}_r \frac{\sqrt{1 + 4\Lambda} - 1}{\Lambda}\right)\right] = 0. \tag{3.43}$$

Coulomb logarithm $\Lambda$ is large for such objects like galaxies, the typical value $\Lambda \sim 200$ and the system of equations (3.42), (3.43) for the large Coulomb logarithm $\Lambda$ simplifies to

$$\mp \frac{1 - r_a^2 k^2}{\sqrt{\pi}} e^{\hat{\omega}'^2 - \hat{\omega}''^2} = \hat{\omega}'' \cos(2\hat{\omega}'\hat{\omega}'') - \hat{\omega}' \sin(2\hat{\omega}'\hat{\omega}''), \tag{3.44}$$

$$\hat{\omega}' \cos(2\hat{\omega}'\hat{\omega}'') + \hat{\omega}'' \sin(2\hat{\omega}'\hat{\omega}'') = 0. \tag{3.45}$$

Let us introduce the notation

$$\alpha = 2\hat{\omega}'\hat{\omega}'', \beta_a = 1 - r_a^2 k^2, \tag{3.46}$$

repeating transformations of Section 2 we obtain the universal equation

$$-e^{\sigma \cot \sigma} \sin \sigma = \frac{\pi}{2\beta_a^2} \sigma, \qquad (3.47)$$

where the notation is introduced $\sigma = -2\alpha = -4\widehat{\omega}'\widehat{\omega}''$. As it was mentioned before, this equation does not depend on the sign in front of parameter $\beta$. The exact solution of equation (121) can be found with the help of the $W$-function of Lambert

$$\sigma_n = \mathrm{Im}\left[W_n\left(\frac{2\beta_a^2}{\pi}\right)\right], \qquad (3.48)$$

frequencies $\widehat{\omega}'_n, \widehat{\omega}''_n$ are (see (2.57))

$$\omega'_n = k\sqrt{-\frac{k_B T}{2m}\sigma_n \tan\frac{\sigma_n}{2}}, \quad \omega''_n = -k\sqrt{-\frac{k_B T}{2m}\sigma_n \cot\frac{\sigma_n}{2}}. \qquad (3.49)$$

In asymptotic for large entire positive $n$ (singular point $r_a k = 1$ is considered further in this section) one obtains

$$\sigma_n = \left(n + \frac{1}{2}\right)\pi, \quad \widehat{\omega}'_n = \frac{\sqrt{\sigma_n}}{2} = \frac{1}{2}\sqrt{\pi\left(n + \frac{1}{2}\right)}, \quad \widehat{\omega}''_n = -\frac{\sqrt{\sigma_n}}{2} = -\frac{1}{2}\sqrt{\pi\left(n + \frac{1}{2}\right)}. \qquad (3.50)$$

The exact solution for the $n$-th discrete solution from the spectrum of oscillations follows from (3.48), (3.49):

$$\widehat{\omega}_n = \frac{1}{2}\sqrt{-\mathrm{Im}\left[W_n\left(\frac{2(1-r_a^2 k^2)^2}{\pi}\right)\right]\tan\left[\frac{1}{2}\mathrm{Im}\left[W_n\left(\frac{2(1-r_a^2 k^2)^2}{\pi}\right)\right]\right]} - \\ \frac{i}{2}\sqrt{-\mathrm{Im}\left[W_n\left(\frac{2(1-r_a^2 k^2)^2}{\pi}\right)\right]\cot\left[\frac{1}{2}\mathrm{Im}\left[W_n\left(\frac{2(1-r_a^2 k^2)^2}{\pi}\right)\right]\right]}. \qquad (3.51)$$

The square of the oscillation frequency of the longitudinal gravitational waves $\widehat{\omega}'^2_n$ is proportional to the wave energy. Hence, the energy of waves is quantized, and as $n$ grows one obtains the asymptotic expression analogous to quantum levels of quantum oscillator in one dimension

$$\widehat{\omega}'^2_n = \frac{\pi}{4}\left(n + \frac{1}{2}\right), \qquad (3.52)$$

the squares of possible dimensionless frequencies become equally spaced as in the plasma case:

$$\widehat{\omega}'^2_{n+1} - \widehat{\omega}'^2_n = \frac{\pi}{4}. \qquad (3.53)$$

or

$$\omega'^2_{n+1} - \omega'^2_n = \frac{\pi}{2} k^2 \frac{k_B T}{m}. \qquad (3.54)$$

This difference can be connected with energy of Newtonian graviton.

From the first glance we could wait for repeating the previous calculations realized for the large Coulomb logarithm $\Lambda$ after application of Eq. (2.55). But it is not so. Introduced value $\beta_a = 1 - r_a^2 k^2$ in comparison with analogical plasma value $\beta = 1 + r_D^2 k^2$ contains the minus signs (not to mention about the different physical sense of $r_a$ and $r_D$) which reflect the difference between Newton and Coulomb laws and leads to fundamental changing in behavior of frequencies in dependence of $r_a k$.

Figures 3.1 and 3.2 reflect the result of calculations for 200 discrete levels for the case of the large Coulomb logarithm $\Lambda$.

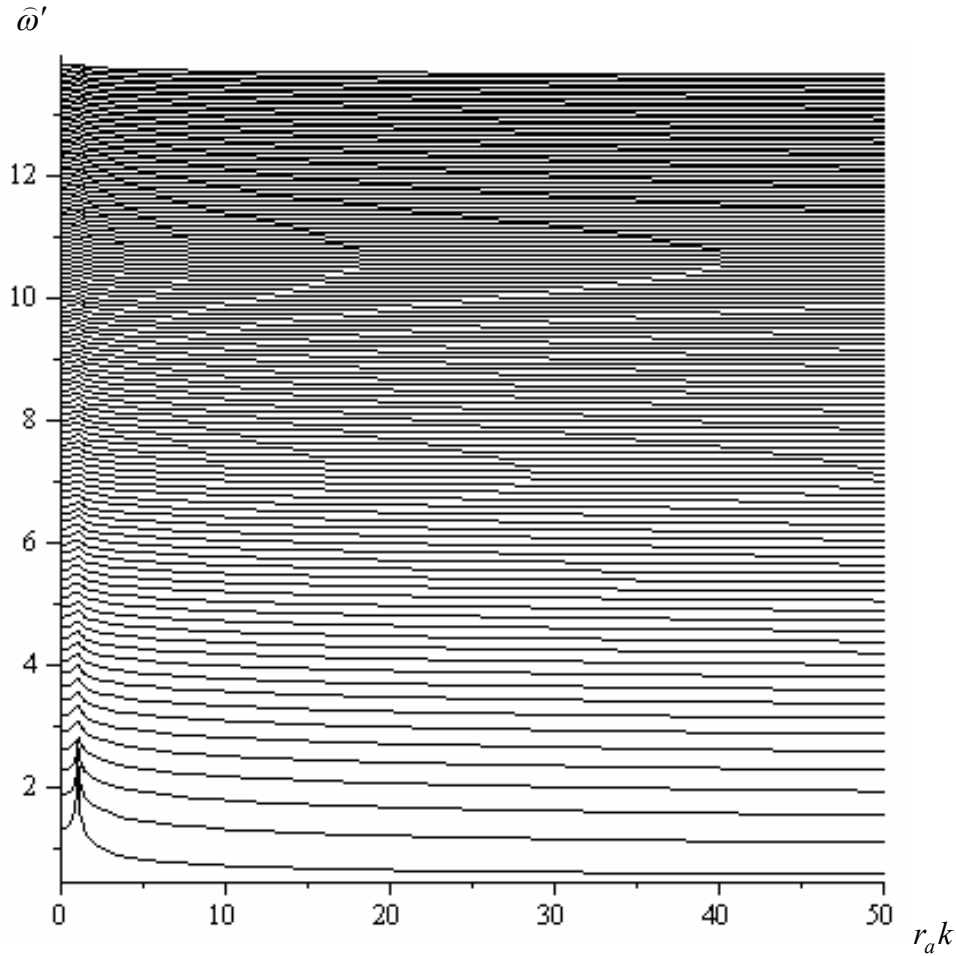

Fig. 3.1. The dimensionless frequency $\hat{\omega}'$ ($y$ axes) versus parameter $r_a k$ ($x$ axes).

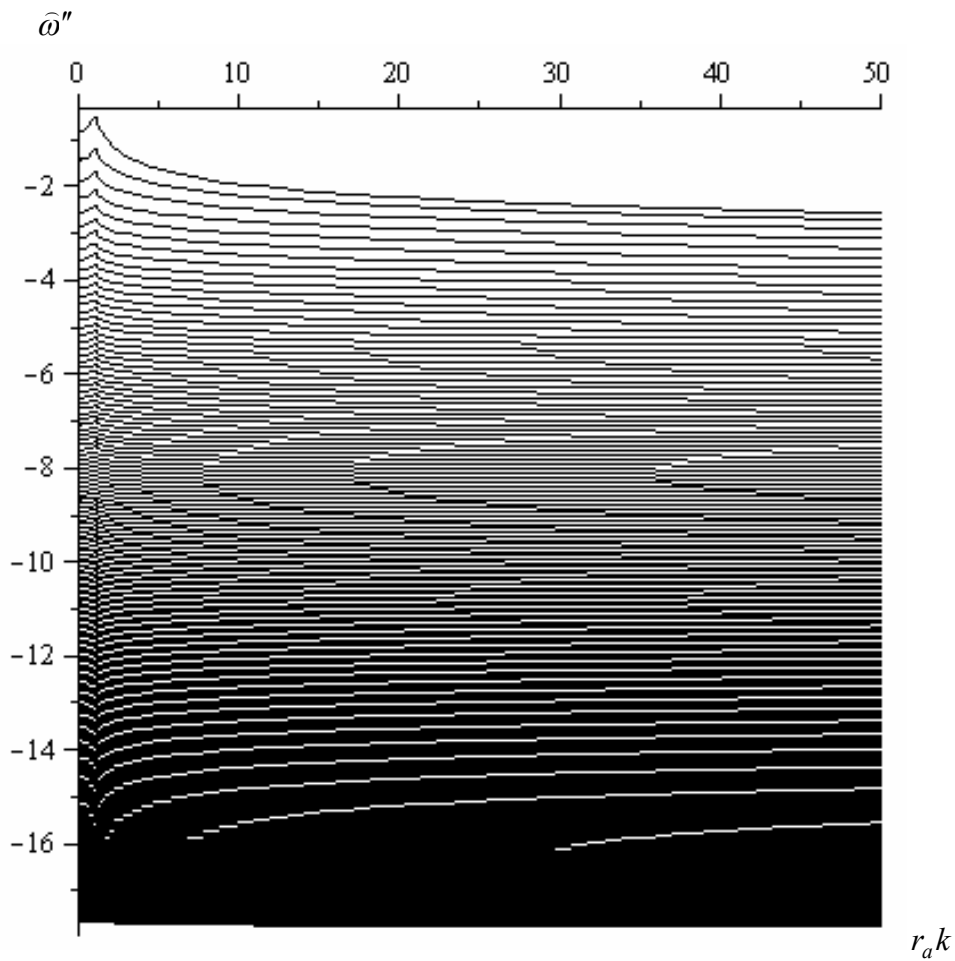

Fig. 3.2. The dimensionless frequency $\widehat{\omega}''$ versus parameter $r_a k$.

Figures 3.3 and 3.4 shows the discrete set of frequencies which vicinity corresponds to passing over from abnormal to normal dispersion for discrete systems in more large interval between 0 and 150 $r_a k$. As in the plasma case these bell like curves have regular character with factor of about two for $r_a k$ corresponding to positions of tops of 'bells'.

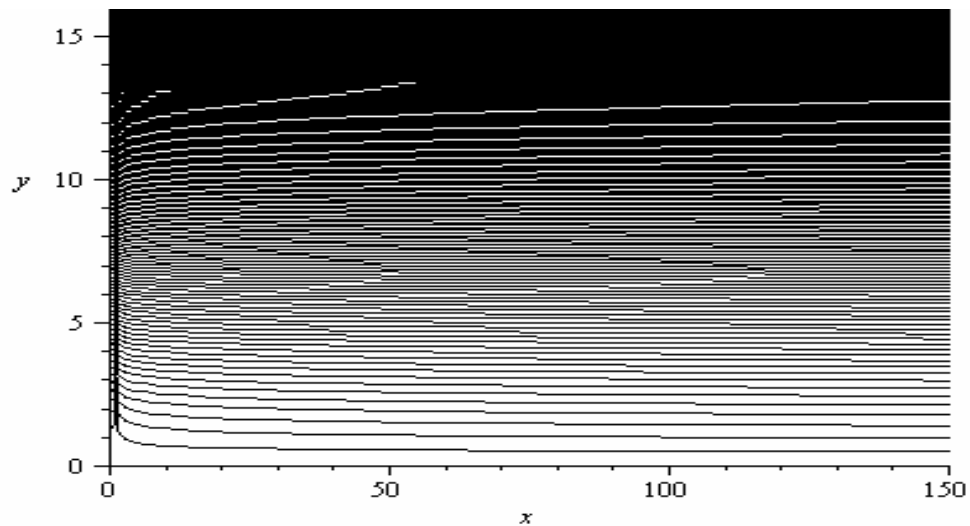

Fig. 3.3. The dimensionless frequency $\widehat{\omega}'$ ($y$ axes) versus parameter $r_a k$ ($x$ axes).

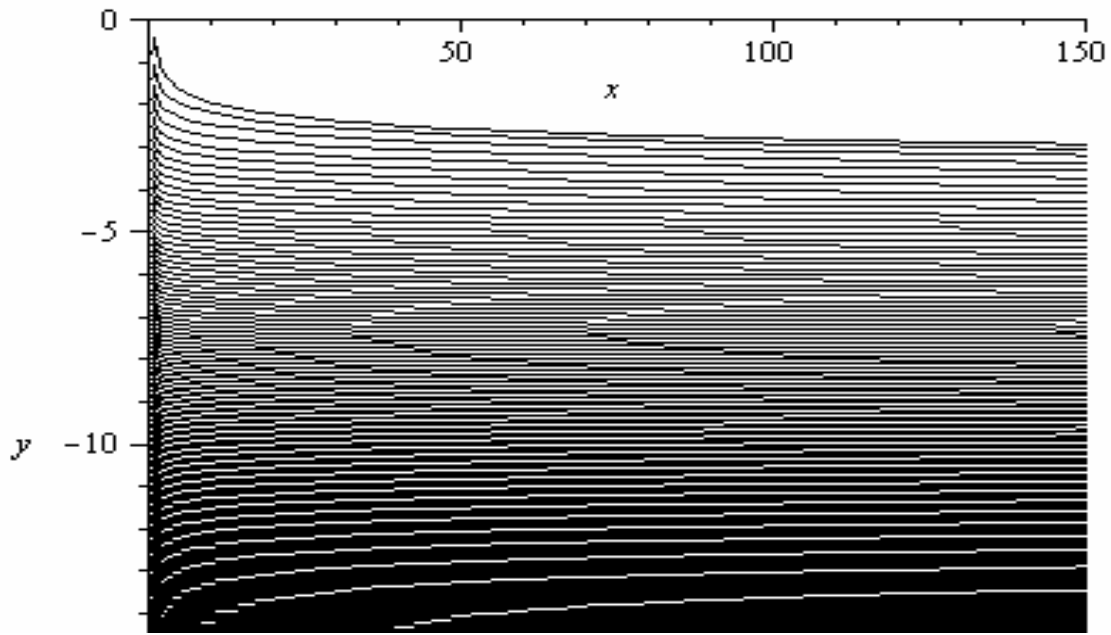

Fig. 3.4. The dimensionless frequency $\widehat{\omega}''$ ($y$ axes) versus parameter $r_a k$ ($x$ axes).

You can see this very complicated topology of curves in Figs. 3.1 – 3.4 including the discrete spectrum of the bell-like curves in the mentioned figures. This singularity was discussed before in the plasma variant of the theory and is connected with the existence of derivatives $d(r_a k)/d\widehat{\omega}'$, $d(r_a k)/d\widehat{\omega}''$. As before this effect has no attitude to the mathematical programming and looks in the definite sense like effect of "shroud of Christ" – self organization of visible information in the human conscience. It seems that the curves of high levels have different topology in comparison with the low levels. But it is far from reality. Look at Figs. 3.5, 3.6 and you see for frequencies $\widehat{\omega}'_{200}$ and for $\widehat{\omega}''_{200}$ the same character features as for lower frequencies.

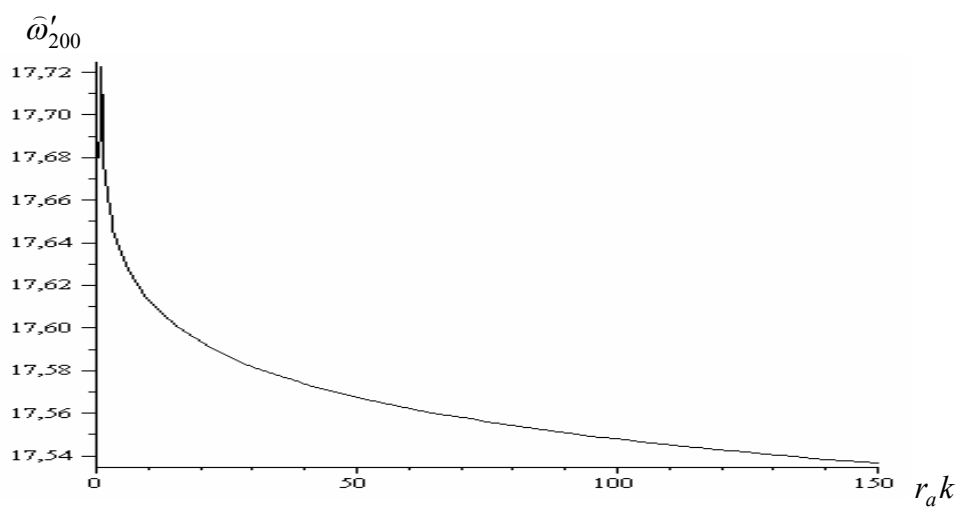

Fig. 3.5. The dimensionless frequency $\widehat{\omega}'_{200}$ versus parameter $r_a k$.

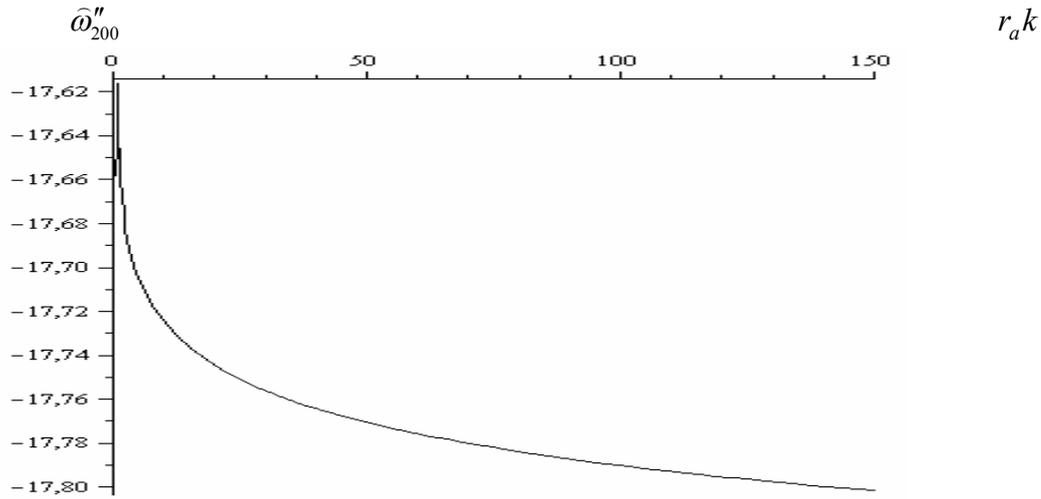

Fig. 3.6. The dimensionless frequency $\widehat{\omega}''_{200}$ versus parameter $r_a k$.

Let us introduce the parameter

$$\xi_a = \frac{\pi}{2\beta_a^2} = \frac{\pi}{2(1 - r_a^2 k^2)^2}. \qquad (3.55)$$

Then Eq. (3.47) can be written as

$$-e^{\sigma \cot \sigma} \sin \sigma = \xi_a \sigma \qquad (3.56)$$

Let us consider the concrete examples of calculations. Suppose that $\xi_a = 0.0002$, it corresponds to $r_a k = 9.47$ or $\lambda = 0.663\, r_a$. The Table 1 contains $\gamma_n$, $\widehat{\omega}'_n$, $\widehat{\omega}''_n$, $n = 1,...,10$; 300, 1000 for $\xi_a = 0.0002$.

Table 1, $\xi_a = 0{,}0002$.

| $n$ | $\gamma_n$ | $\widehat{\omega}'_n$ | $\widehat{\omega}''_n$ |
|---|---|---|---|
| 1 | 1.714 $\pi$ | 0.806 | -1.671 |
| 2 | 3.602 $\pi$ | 1.430 | -1.978 |
| 3 | 5.560 $\pi$ | 1.900 | -2.298 |
| 4 | 7.541 $\pi$ | 2.283 | -2.594 |
| 5 | 9.530 $\pi$ | 2.610 | -2.867 |
| 6 | 11.523 $\pi$ | 2.901 | -3.119 |
| 7 | 13.518 $\pi$ | 3.165 | -3.354 |
| 8 | 15.515 $\pi$ | 3.408 | -3.575 |
| 9 | 17.513 $\pi$ | 3.635 | -3.784 |
| 10 | 19.511 $\pi$ | 3.848 | -3.982 |
| 300 | 599.500 | 21.707 | -21.691 |
| 1000 | 1999.500 | 39.636 | -39.620 |

It is of interest to investigate the singular point where

$$r_a k = 1 \qquad (3.57)$$

and $\xi_a$ turns into infinity. Note that when $\xi_a \to \infty$, in the vicinity of $r_a k = 1$ the solution $\sigma \to \pi + 0$ and therefore $\bar{\omega}' \to \infty$ and $\bar{\omega}'' \to 0$. But phase velocity of wave $u_\phi = \omega' r_a$ and phase velocity of gravitational wave turns into infinity and damping is equal to zero. In vicinity of $r_a k = 1$ one obtains "gravitational window" with increasing of frequency $\bar{\omega}'_n$ and decreasing of damping; the corresponding wave lengths $\lambda_a$ is

$$\lambda_a = 2\pi r_a. \qquad (3.58)$$

Let us make some estimations. The mean density (calculated via luminous matter in stars and galaxies) $\rho \approx (0.01 - 0.02)\rho_c$, where $\rho_c = 0.47 \cdot 10^{-29}$ $g/cm^3$. Using also for estimation $T = 3K$, $n = 0.3 \cdot 10^{-7}$ $cm^{-3}$, $m = 1.6 \cdot 10^{-24} g$, one obtains from relation (3.31) $r_a = 0.82 \cdot 10^{23} cm = 0.027 Mpc$ and $\lambda_a = 0.17 Mpc$.

Now we can create the physical picture leading to the Hubble flow.

*The main origin of Hubble effect (including the matter expansion with acceleration) is self – catching of expanding matter by the self – consistent gravitational field in conditions of weak influence of the central massive bodies.*

The relation (3.58) is the condition of this self – catching as result of explosion with appearance of waves for which the wave lengths is of about $\lambda_a$. Gravitational self – catching takes place for Big Bang having given birth to the global expansion of Universe, but also for Little Bang [13] in so called Local Group of galaxies. The gravitationally bound system of the Local Group can exist only within region $r < r_v$, where gravitation dominates of central mass. Outside the Group at distances $r > r_v$ the Hubble flow of Galaxies starts. This "no reentering radius" was found as result of direct observations of the Local Group: $r_v \sim 1 Mpc$. This value is in good coincidence with the lengths $\lambda_a \sim 0.2 Mpc$.

Some important remarks should be done:

1. Effects of gravitational self-catching should be typical for Universe. The existence of "Hubble boxes" is discussed in review [13] as typical blocks of the nearby Universe.

2. As it follows from Figs. 3.1 – 3.6 effect of gravitational self-catching exists *in finite region* close to $r_a k = 1$, the phase velocity is defined by discrete spectrum $u_{\phi,n} = \bar{\omega}'_n \sqrt{\dfrac{2k_B T}{m}}$.

3. Gravitational self-catching can be significant in the Earth conditions.

The last remark needs to be explained. Gravitational self-catching can be essential in the physical systems which character lengths correlates with $r_a$. For water by the earth conditions $T = 300$ $K$, $\rho = 1 g/cm^3$; $\gamma_N = 6.6 \cdot 10^{-8} cm^3/(g \cdot s^2)$, the lengths $r_a = 407.43 km$ and $\lambda_a = 2558.66 km$. For close collisions $r_c \sim 10^{-7}$ cm and Coulomb logarithm $\Lambda_a = ln\dfrac{r_a}{r_c} \sim 10^2$.

The delivered theory can be applied in the Earth conditions if the influence of central mass can be excluded from consideration. This condition realizes in the tsunami motion because the direction to the Earth center supposes perpendicular to the direction of additional self-consistent gravitational force acting in the tsunami wave. In essence, the catching of water mass is realizing by longitudinal self-consistent gravitational wave. I don't intend here to go into details, but the origin of effects of the small attenuation can be qualitatively explained from position of kinetic theory.

Let us calculate the mean velocity $\bar{u}_+$ of particles moving in a chosen direction. If this direction is considering as the positive ones, then $u > 0$ and for the Maxwellian function $f_0$

$$\bar{u}_+ = \sqrt{\frac{m}{2\pi k_B T}} \int_0^{+\infty} e^{-mu^2/2k_B T} u\, du = \sqrt{\frac{k_B T}{2\pi m}}. \qquad (3.59)$$

Kinetic energy, connected with this motion is

$$\frac{m\bar{u}_+^2}{2} = \frac{k_B T}{4\pi}. \qquad (3.60)$$

From relation (3.31) follows

$$\gamma_N m \rho r_a^2 = \frac{k_B T}{4\pi} \qquad (3.61)$$

or

$$\frac{m\bar{u}_+^2}{2} = \gamma_N m \rho r_a^2, \qquad (3.62)$$

$$\bar{u}_+ = r_a \sqrt{2\rho \gamma_N} \qquad (3.63)$$

Therefore if the selected direction is opposite to the direction of the wave motion, energy of gravitational field $E_a = \gamma_N m \rho r_a^2$ (per particle) should be applied for exclusion of such kind of particles. For water in considered estimation one obtains $E_a = \gamma_N m \rho r_a^2 = 3.293 \cdot 10^{-15} erg$, $\bar{u}_+ = r_a \sqrt{2\rho \gamma_N} = 533 km/hour$ - the typical value of tsunami in ocean. Otherwise the wave expansion leads to the energy dissipation of directional motion at the form of the chaotic heat motion. But in the case if the forces of gravitation attraction liquidize (or keep to a minimum) these losses, the wave is moving without attenuation.

**Conclusion**

Landau rule for the analytical estimation of the corresponding singular integral in plasma physics should be considered as additional condition imposed on the physical system. Another condition with the transparent physical sense (connected with the calculation of independent oscillations in collisional physical system) leads to the appearance of spectrum of oscillations. The Landau formulation contains two restrictions of principal significance – potential of the force field and maxwellian function as approximation for distribution function. Then the modified Landau formulation can be imposed in consideration of other physical systems, particularly in collisional damping of gravitational waves in the Newtonian matter. The generalized theory of Landau damping is applied to the gravitational physical systems in the framework of plasma – gravitational analogy. The exact solution of the corresponding dispersion equations is obtained. The results of calculations lead to existence of the discrete spectrum of frequencies and the discrete spectrum of dispersion curves. The existence of gravitational window is discovered where the waves of gravitating matter are expanding without damping.

Moreover the aim of investigations undertaken here and in [5 – 8] consists in creation of the unified non-local quantum theory of transport processes. It is not a new idea that physics is unified construction but not the collection of inconsistent facts. I hope this paper demonstrate the validity of this conception.

# REFERENCES

1. Landau L.D. *J. Phys. USSR* **10** 25 (1946).
2. Landau L.D. *Zh. Eksp. Teor. Fiz.* **16** 574 (1946).
3. Dawson J. Phys. Fluids **4** 869 (1961).
4. Soshnikov V.N. Some paradoxes of the Landau damping. Supplement to the Russian translation of the book: Clemmow P.C., Dougherty J.P. Electrodynamics of Particles and Plasmas, Reading, Mass.: *Addison-Wesley Publ.* Co. (1990).
5. Alexeev B.V., J. Nanoelectron. Optoelectron. 4, 186 - 199 (2009).
6. Alexeev B.V., J. Nanoelectron. Optoelectron. 3, 143 - 158 (2008).
7. Alexeev B.V., J. Nanoelectron. Optoelectron. 3, 316 - 328 (2008).
8. Alexeev B.V. Generalized Boltzmann Physical; Kinetics. Elsevier, Amsterdam, The Netherlands (2004).
9. Chapman S., T.G. Cowling T.G., "The Mathematical Theory of Non-uniform Gases", Cambridge: At the University Press, (1952)
10. Klimontovich Yu.L. About Necessity and Possibility of Unified Description of Hydrodynamic Processes. *Theoretical and Math. Physics*, **92** (2) p.312 (1992).
11. Bell J.S. On the Einstein Podolsky Rosen Paradox. *Physics*, v.1, p. 195 (1964).
12. Shakhov E.M. Method of Rarefied Gas Investigation. (M.: *Nauka*, (1974).
13. Chernin A.D. Physics – Uspekhi, 51 (3), 267 – 300 (2008).
14. Gliner E.B. Sov. Phys. Dokl. 15, 559 (1970).